# Superionic states formation in group III oxides irradiated with ultrafast lasers

R.A. Voronkov[1*], N. Medvedev[2,3], A.E. Volkov[1,]

[1]P. N. Lebedev Physical Institute of the Russian Academy of Sciences, Leninskij pr., 53,119991 Moscow, Russia;

[2]Institute of Physics, Czech Academy of Sciences, Na Slovance 2, 182 21 Prague 8, Czech Republic;

[3]Institute of Plasma Physics, Czech Academy of Sciences, Za Slovankou 3, 182 00 Prague 8, Czech Republic;

## Abstract

A number of group III-metal oxides are studied *via* density functional theory in order to establish a possibility of nonthermal transition of these materials into a superionic state. Atomic and electronic properties of the materials are analyzed during the transitions to acquire insights into physical mechanisms guiding such transformations. This knowledge is then used to suggest a criterion allowing to predict the possibility of such transitions without employing computationally heavy methods.

## 1. Introduction

Two main mechanisms govern structure transformations in materials excited with a high-intensity femtosecond laser pulses [1]. A thermal one realizes via electron-atom (electron-phonon) coupling equilibrating a highly excited electronic system with the ionic one [2]. This coupling results in a noticeable lattice heating when an amount of energy transferred from electrons to atoms becomes significant, typically at timespans > 1 ps [2,3].

The second one is structural transitions via nonthermal channel, which may occur at much shorter times [4–7]. Nonthermal transition are caused by a significant energy density deposited into the electronic system, usually corresponding to the electronic temperature of a few eV [8]. Such excitation of the electronic system temporarily changes a potential energy surface of the atomic system in a solid. Appearing uncompensated forces initiate movements of atoms trying to find their new equilibrium positions. In some materials, depending on the excitation level, these new equilibrium positions may not exist, and nonthermal melting ensues even when the atomic temperature does not exceed the melting point of the material [6,7,9,10].

Nonthermal atomic movement dominates at timescales under 0.5-1 ps, until the thermal channel starts to play a role [1]. It may lead to exotic structures and phases (usually transient) that are not achievable at equilibrium conditions [8,11]. One of such unusual phases is superionic state, which consists of one subsystem of a compound in a liquid state, whereas the other one is in a solid phase, simultaneously. It was experimentally confirmed that a superionic state does exist. It was recently produced by laser-induced dynamical shock compression of water ice [15]. Presumably, superionic ice can be found naturally in giant planets interiors such as Uranus, Neptune, or other exoplanets [16].

Superionic alumina was predicted to form after fs-laser irradiation [11]. This state with liquid oxygen sublattice and solid aluminum one stable during hundreds of femtoseconds may be produced at excitation levels achievable in irradiation spots of modern free-electron lasers [12–14].

However, general mechanisms leading to superionic state formation in materials were not discussed in Ref. [11] because only $\alpha$-$Al_2O_3$ (corundum) was investigated. In this paper, we study a possibility of nonthermal transitions in a few oxides formed from group III metal: $Al_2O_3$, $Ga_2O_3$, $In_2O_3$, and $In_2S_3$ for comparison, to gather some statistics of materials demonstrating superionic behavior under extreme electronic excitation. These materials also have the same R-3c group symmetry as that of $\alpha$-$Al_2O_3$ and their electronic structure is formed by overlap of $ns^2np^1$ and $ms^2mp^4$ atomic orbitals. Additionally, we check polymorphs of these materials for the existence of a superionic state. By analyzing such material properties as group symmetry, chemical composition and electronic energy levels, we investigate which of the parameters may be the driving force of the transition into the superionic state.

We identify thresholds of electronic excitation required to produce the superionic states in some of these compounds. Threshold doses obtained in this work should be a reasonable reference point for future experiments.

After that, we analyze similarities in fundamental properties of these materials to find criteria that would allow predicting a possibility of superionic state creation without carrying out computational resource-intensive *ab-initio* molecular dynamics simulations.

## 2. Methods

For all presented simulations, we use the density functional molecular dynamics simulation within the Quantum Espresso simulation package [17]. We use norm-conserving pseudopotentials from the Quantum Espresso library and Perdew-Burke-Ernzerhof (PBE) exchange-correlation functional [18]. Although non-hybrid functionals are known to underestimate the band gap value at ambient conditions, in a case of high electronic temperatures they perform much better [19].

To study atomic and electronic structure dynamics during nonthermal transitions we implement the following algorithm. After the standard procedure of a geometry optimization, the lattice is allowed to thermalize at the room temperature $T_i$ = 300 K via density functional theory (DFT) molecular dynamics with unperturbed electronic system ($T_e$ = 300 K set via Fermi-Dirac smearing) during 500 fs with 0.5 fs time step. After that, a molecular dynamics simulation within 1000 fs with 0.5 fs time step is performed with electronic temperature instantly elevated to a certain value assuming that electrons adhere to the Fermi-Dirac distribution.

Within this procedure, we assume that electron cascades after laser irradiation can be neglected since they take a few femtoseconds in a typical FEL spot except for hard X-rays and does not significantly affect lattice dynamics [20]. This allows us to instantly apply thermalized distribution for the electronic ensemble avoiding complex consideration of a short-living nonequilibrium stage that cannot be treated in DFT.

We further assume that the electronic temperature can stay constant throughout the simulation since electron energy loss *via* kinetic energy transfer to the lattice or *via* spatial dissipation is minor within ~500 fs in the central area of the irradiated volume. These assumptions are supported by the nonadiabatic tight-binding MD simulations in Ref. [1]. We note, however, that all simulations beyond 500 fs in this paper are performed only in order to confirm nonthermal transition behavior when it is unclear at shorter timescales for some compounds investigated here.

We do not consider excitonic effects as well, since we assume that a significant level of electronic excitation and induced atomic perturbations studied here do not allow excitons to form.

An NVT-ensemble (constant number of particles, volume and temperature) is used for the electronic system and an NVE-ensemble (constant number of particles, volume and energy) for the atomic system. This choice corresponds to the conditions in the bulk achieved after irradiation with an FEL pulse, where the unperturbed media maintains a constant volume of the target's excited part in the bulk for times sufficiently longer than those modeled here [21].

Apart from the DFT-MD simulations, further analyses were performed. In order to identify affiliation of obtained energy levels to atomic orbitals, a dependence of gamma-point energy levels on the interatomic distance was constructed *via* Parrinello-Raman variable-cell molecular dynamics [22] with a target pressure $P_{target}$ = -600 kbar and atoms kept in the ideal lattice positions. For this analysis, the initial state of a material was set to its ambient structure and zero atomic and electronic temperatures (the latter was set via Fermi-Dirac distribution at $T_e \approx 30$ K). At each molecular dynamics step, energy levels were extracted and shifted to zero chemical potential (Fermi level). At the last step, the gamma-point projected electronic density of states (PDOS) was calculated.

Atomic potential energy surfaces were constructed in a series of calculations. For example, for one Ga or one O atom (in a fixed lattice of all other atoms) a uniform 3-dimentional mesh was set on Cartesian coordinates from ($a_0$-1) Å to ($a_0$+1) Å, where $a_0$ is a coordinate ($x,y$ or $z$) of the equilibrium position, with 0.1 Å step. For each point of the mesh, a self-consistent calculation was carried out. Then, the total energy of the electronic system representing atomic potential energy was calculated (excluding pseudo-nuclei repulsion term that becomes noticeable only at much shorter interatomic distances than those considered here).

Energy cutoff parameter was set to $E_{cut} \approx 952$ eV (70 Ry) for nonthermal transitions simulations, potential energy surfaces construction and PDOS calculations for materials at ambient conditions. For energy levels of the expanding cell and corresponding PDOS calculations, the parameter was set to $E_{cut} \approx 1360$ eV (100 Ry) since increasing interatomic distance requires more plane waves in the DFT basis set to describe electronic states becoming more localized around nuclei. For all calculations, supercells consisting of 80 atoms were used.

## 3. Results

### 3.1. Atomic properties.

Calculated thresholds for nonthermal phase transitions in studied materials are presented in Table 1. A threshold of the absorbed dose triggering nonthermal transition, its equivalents in electronic temperature and number of valence electrons excited to the conduction band, initial structure as well as type of the transition (superionic or melting) are shown for each material.

**Table 1**. Threshold parameters triggering nonthermal transitions in various materials in various phases. Here, $T_e$ is the electronic temperature, $D$ is the absorbed dose, and $N_e$ is the percentage of valence electrons excited to the conduction band.

| Material | Phase | Type of the transition | $T_e$, eV | $D$, eV/atom | $N_e$, % |
|---|---|---|---|---|---|
| $Al_2O_3$ | α, R-3c | superionic | 2.75 | 1.4 | 4.8 |
| | Ia-3[a] | superionic | 2.75 | 1.64 | 5.7 |
| $Ga_2O_3$ | α, R-3c | superionic | 2.25 | 1.2 | 4.9 |
| | β, C2/m | melting | 1.75 | 0.6 | 3.4 |
| | δ, Ia-3 | superionic | 2.25 | 1.3 | 5.3 |
| $In_2O_3$ | rh, R-3c | superionic | 2 | 1.2 | 5.3 |
| | c, Ia-3 | superionic | 2 | 1.2 | 5.4 |
| $In_2S_3$ | ε, R-3c | melting | 1.25 | 0.8 | 5.5 |

For all materials becoming superionic during nonthermal transition (except for c-$In_2O_3$ discussed below) we observed the same profile of mean atomic displacements. During the first ~250-300 fs, both metal (Me) and O atoms are rapidly moving away from their ambient equilibrium positions. After that, displacements of Me atoms saturate at values ~0.6 Å, while O atoms keep moving demonstrating diffusive (liquid-like) behavior. One can see an example of typical displacements in Figure 1. The detailed information about each transition can be found in supplementary materials.

---

[a] To the best of our knowledge, Ia-3 phase of $Al_2O_3$ appears only in simulations [30]. It was studied here for the sake of comparison of different materials with the same space group symmetry.

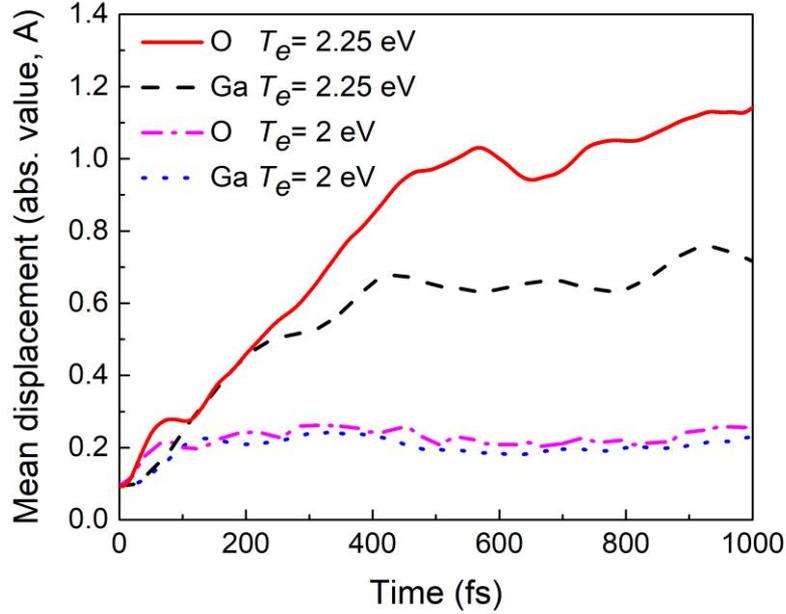

**Figure 1.** Mean atomic displacements in Ia-3 phase of $Ga_2O_3$ at the electronic temperature (2 eV) below the superionic threshold and at the threshold electronic temperature (2.25 eV).

We note that for Ia-3 phase of $In_2O_3$, nonthermal transition to the superionic state at $T_e = 2$ eV can be clearly identified only at times >1 ps which is far longer than times at which the nonthermal transition channel can exist without noticeable interference from the thermal one; we thus expect that it may not be observable in experiments at the threshold dose. However, increase of the electronic temperature to $T_e = 2.25$ eV results in much faster transition within <500 fs, which should be observable.

As can be seen from Table 1, the space group is not a main parameter driving the transition to a superionic state. $In_2S_3$ nonthermally melts, while other compounds with the same space group become superionic. $Y_2O_3$ (Ia-3 space group) in our previous study also did not demonstrate existence of a superionic state [4].

A mere presence of oxygen atoms in a structure does not guarantee appearance of the superionic state, since C2/m phase of $Ga_2O_3$ exhibits melting, while all other studied materials with oxygen turn into superionic state.

Nevertheless, Table 1 confirms the idea that nearly all group III metal oxides can exhibit superionic behavior after ultrafast sufficient electronic excitation, although some irregularities occur in certain materials or their polymorphs. Thus, it seems that the origin of the ability to transform into a superionic material does not lie in the plane of simple properties such as atomic structure and chemical composition, and an analysis of electronic properties is required.

### 3.2. Electronic properties.

We compared calculated projected electronic density of states (PDOS) of each ambient material from Table 1 (all these PDOS can be found in supplementary materials). Although all PDOS have some similarities and differences, there does not seem to be a definitive characteristic feature that differentiate C2/m phase of $Ga_2O_3$ and $In_2S_3$ from other materials and that could be interpreted as an unambiguous indicator of an ability or inability of a material to exhibit transition to the superionic state. This suggests that electronic energy levels structure itself does not determine a type of a transition. Instead, a complex interplay of various material properties affects formation of the superionic state.

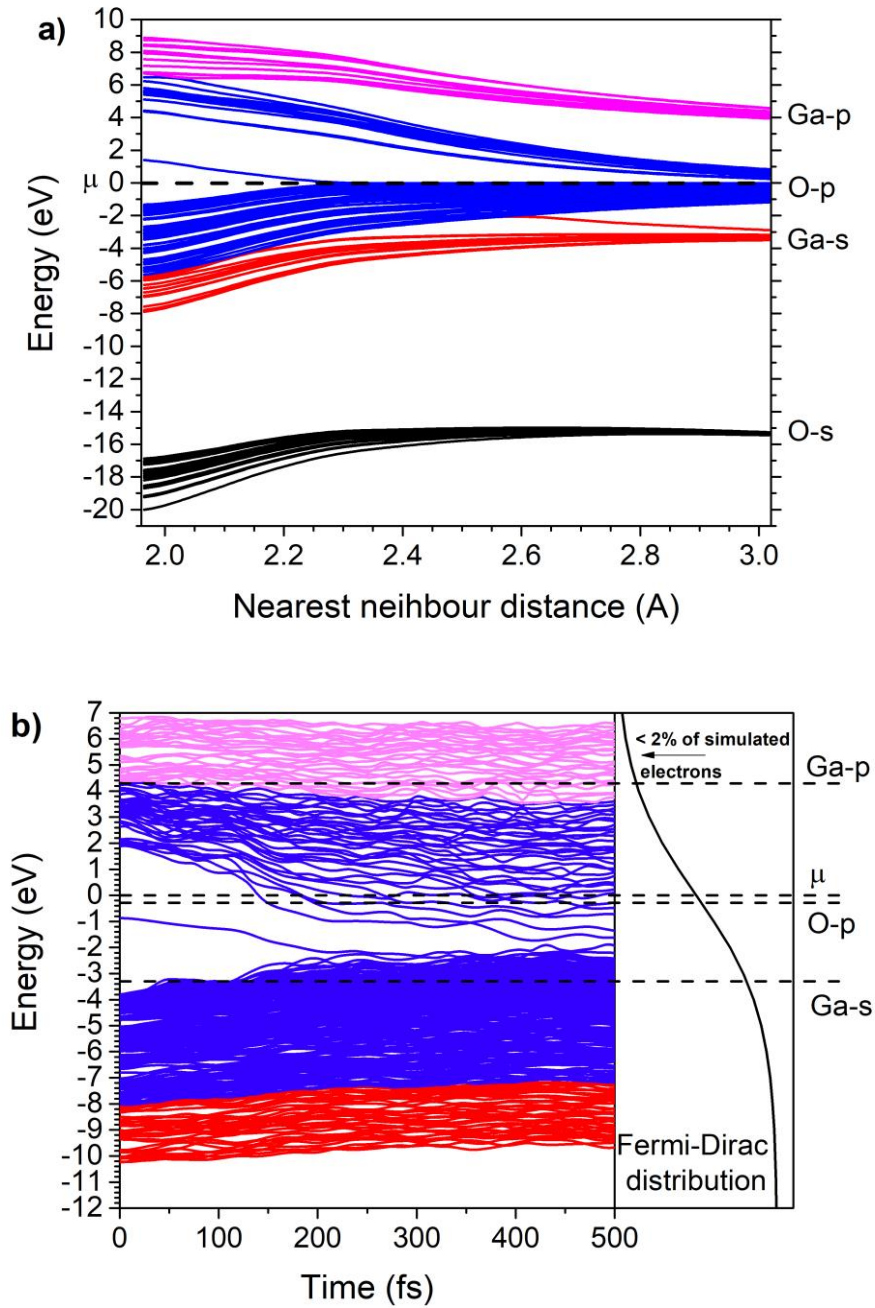

**Figure 2**. (a) Dependence of gamma-point energy levels of Ia-3 phase of Ga$_2$O$_3$ on interatomic distance; (b) Gamma-point energy levels dynamics during nonthermal transition into the superionic state. Here, μ is the chemical potential that is set to zero.

Figure 2a demonstrates a dependence of gamma-point energy levels in Ia-3 phase of Ga$_2$O$_3$ on the interatomic distance. From it, one can see that the upper energy levels of the valence band and the lower levels of the conduction band are formed from oxygen $p$-orbitals. This means that at elevated electronic temperatures triggering the transition into the superionic state, electrons are mainly excited from oxygen bonding energy levels that are shifted below those of isolated atom to oxygen antibonding (shifted above) levels while occupation numbers of metal bonding and antibonding levels remain almost untouched. Moreover, electronic temperature and atomic movement cause these levels to shift (Fig. 2b) making some of the $p$-levels of Ga energy levels (marked as "Ga-$p$") to become bonding. By the end of the simulation (500 fs) only ~1.8% of simulated electrons are occupying "Ga-$p$" antibonding levels. All this indicates that the electronic excitation is affecting potentials in a way that may lead to oxygen melting while preserving metallic sublattice.

Further electronic temperature increase should involve more Ga energy levels both in valence and conduction bands resulting in nonthermal melting of the whole material or at least significant changes in the Ga sublattice. This is exactly what happened with alumina in our previous study [11].

Energy levels behavior from Figure 2a is qualitatively the same for all materials and phases from the Table 1. This means that C2/m phase of $Ga_2O_3$ as well as $In_2S_3$ should also have been transformed into a superionic state if the above reasoning is sufficient. Considering that these phases demonstrate melting instead of the superionic behavior, it means more than one material property affects the superionic state formation.

### 3.3. Potential energy surfaces.

Analyzing properties of materials from Table 1, one finds that ε-$In_2S_3$ is a very unstable material that is starting to turn into β-$In_2S_3$ already at 40ºC [23]. This means that atoms in ε-$In_2S_3$ are in a very shallow potential well. In such a case, almost any significant external perturbation would lead to destabilization of the material.

In contrast, C2/m is the most stable phase of $Ga_2O_3$, while Ia-3 phase is more "exotic" and less stable at ambient conditions (it turns into ε-$Ga_2O_3$ at >500 °C) [24]. Nevertheless, C2/m phase nonthermally melts in contrast to Ia-3 phase exhibiting solid-superionic phase transition. This difference may arise because of asymmetry in Ga potential surface that may result in preferential direction for atomic movement. In combination with thermal oscillations and interatomic potential changes after laser irradiation, this may be a source of easier destabilization of Ga lattice and consequently may lead to melting instead of a transition into superionic state after the electronic temperature elevation.

To illustrate this difference in $Ga_2O_3$ phases, we constructed potential energy surfaces of Ga atoms for Ia-3 and C2/m polymorphs (Figure 3).

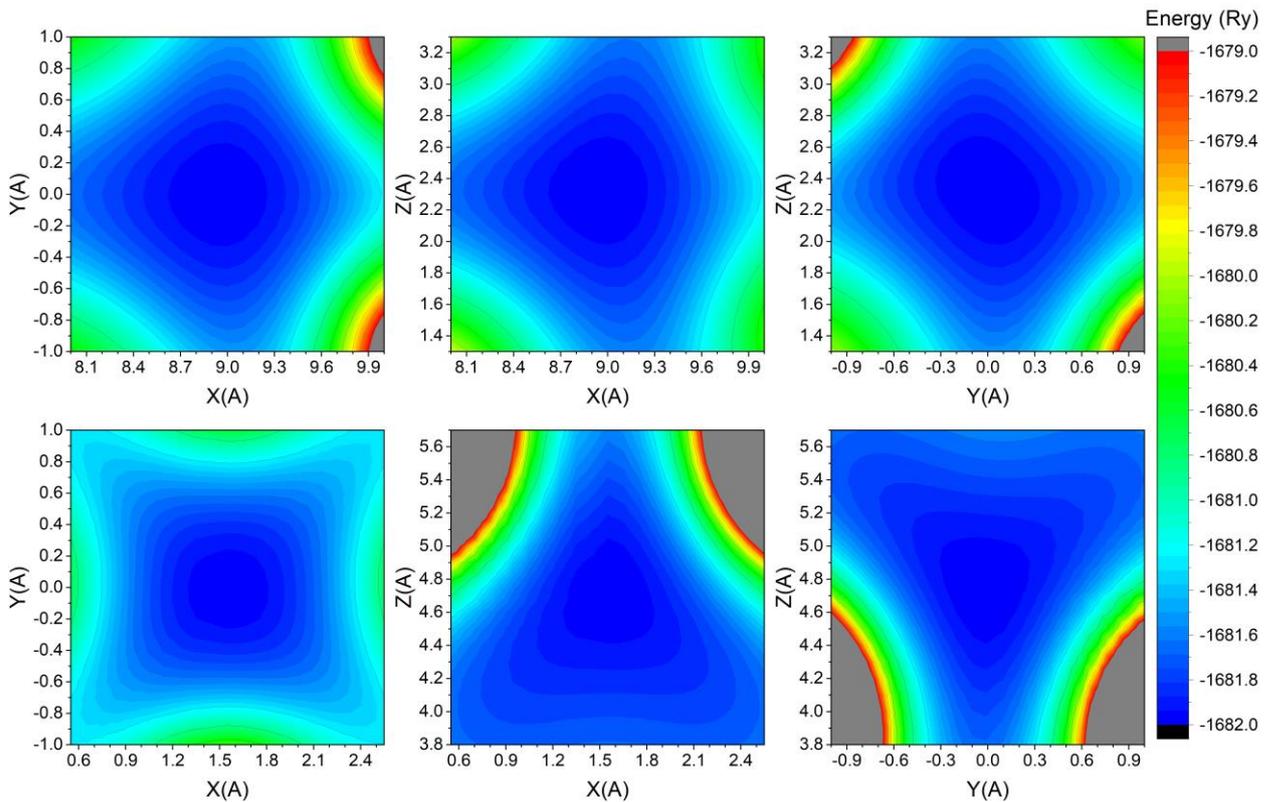

**Figure 3**. Potential energy surfaces of Ga atoms in Ia-3 phase (upper row) and C2/m phase (lower row) projected onto different planes of $Ga_2O_3$.

Figure 3 shows that indeed in C2/m phase, the potential energy surface of Ga is much more asymmetric and has highly distinctive preferential directions. This asymmetry also could be suspected from analysis of the macroscopic material properties: $\beta$-$Ga_2O_3$ has a significantly different thermal conductivity along different axes [25].

Figure 4 shows that Ga atoms in C2/m phase mainly move along $x$ axis during the first ~200-250 fs with displacements twice as large as those along $y$ and $z$ axes. These displacements, however, are significantly smaller than displacements of metal atoms for any other material considered in this work or in our previous papers [4,11]. This also indicates that C2/m phase of $Ga_2O_3$ is an anomaly in the sense of nonthermal transitions.

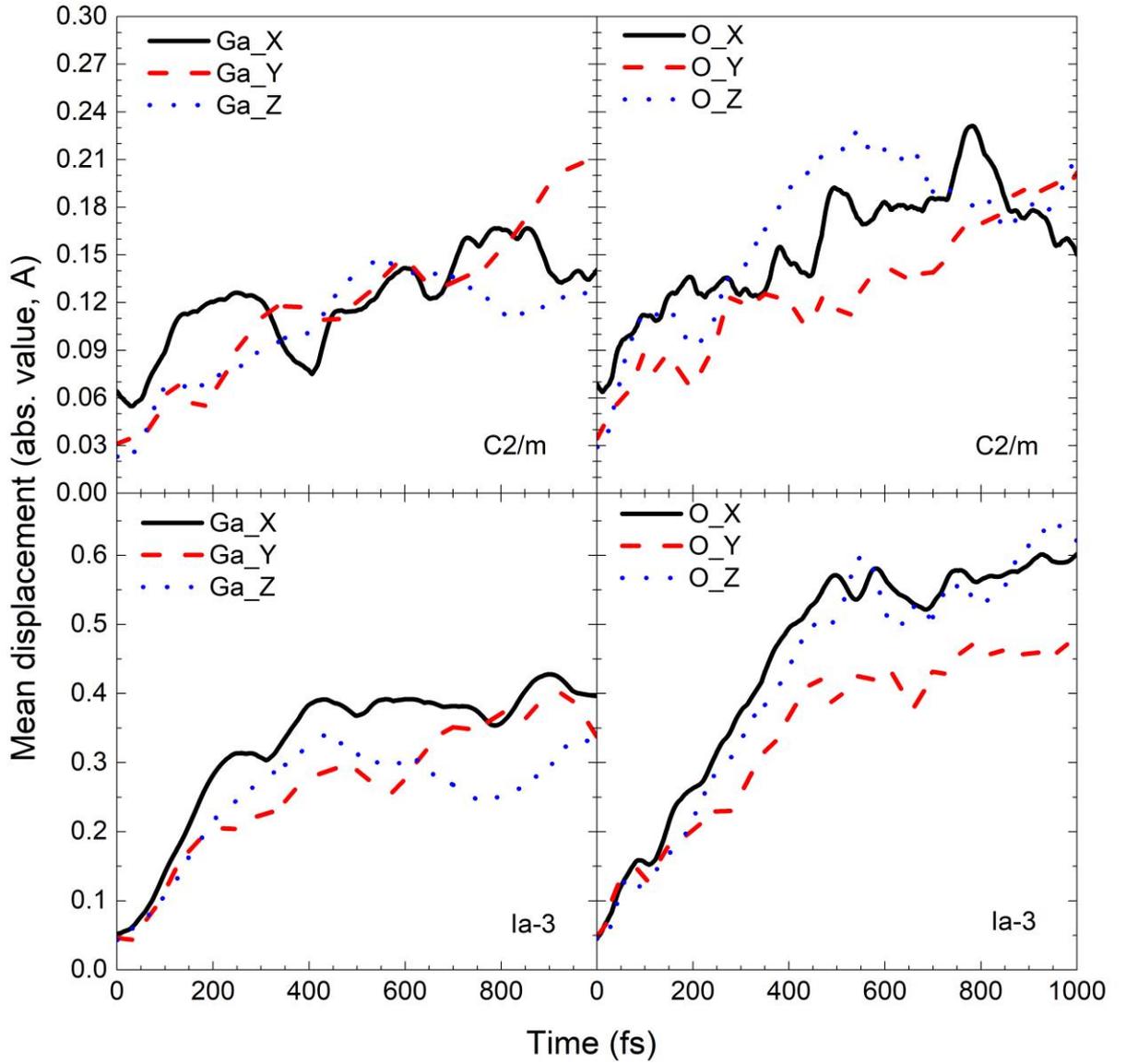

**Figure 4.** Mean atomic displacements along lattice vectors in C2/m phase (top) and Ia-3 phase of $Ga_2O_3$ (bottom) at threshold temperatures.

## 4. Discussion

To elucidate a path of the nonthermal transition into a superionic state, we constructed potential energy surfaces of Ga and O atoms in Ia-3 phase at ambient conditions and at elevated electronic temperature at the initial moment of the transition (see Figure 5).

As mentioned above, Ga atoms lie at a symmetric potential energy surface. At the same time, oxygen potential energy surface has clearly visible preferential directions along Y and X axes. This is confirmed by the mean displacements of oxygen presented in Figure 4: within the first ~100 fs oxygen moves primarily along X and Y axes until potential energy surface profile will change and movement along X and Z axes will become dominant.

Also energy isosurfaces (constant energy surfaces) in all pictures are wider for O atoms. This means that the oxygen potential well is slightly shallower than the gallium one allowing for easier displacements of oxygen atoms.

Therefore, an asymmetry of the potential energy surface of one of the atomic species may serve as an indicator of a material ability to form superionic states under extreme electronic excitation.

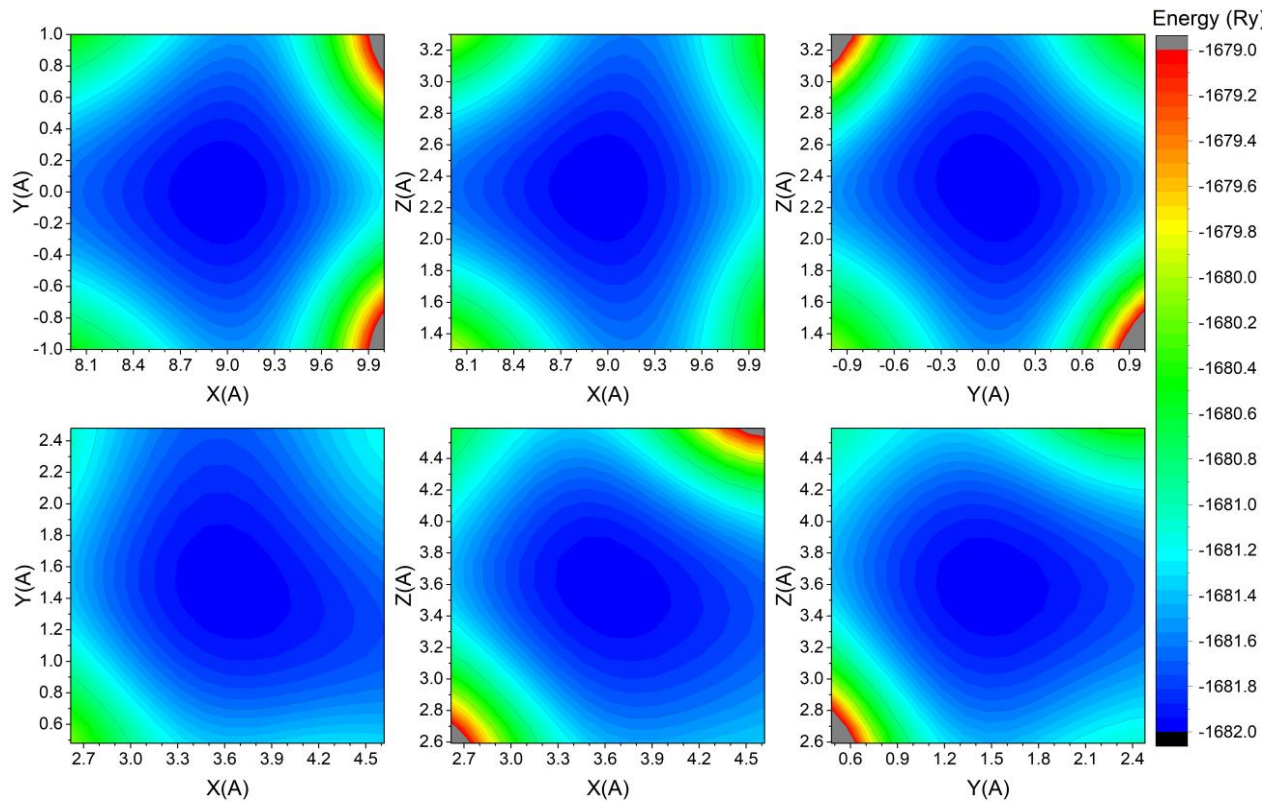

**Figure 5.** Potential energy surfaces of Ga (upper row) and O (lower row) atoms in Ia-3 phase of $Ga_2O_3$ at ambient conditions.

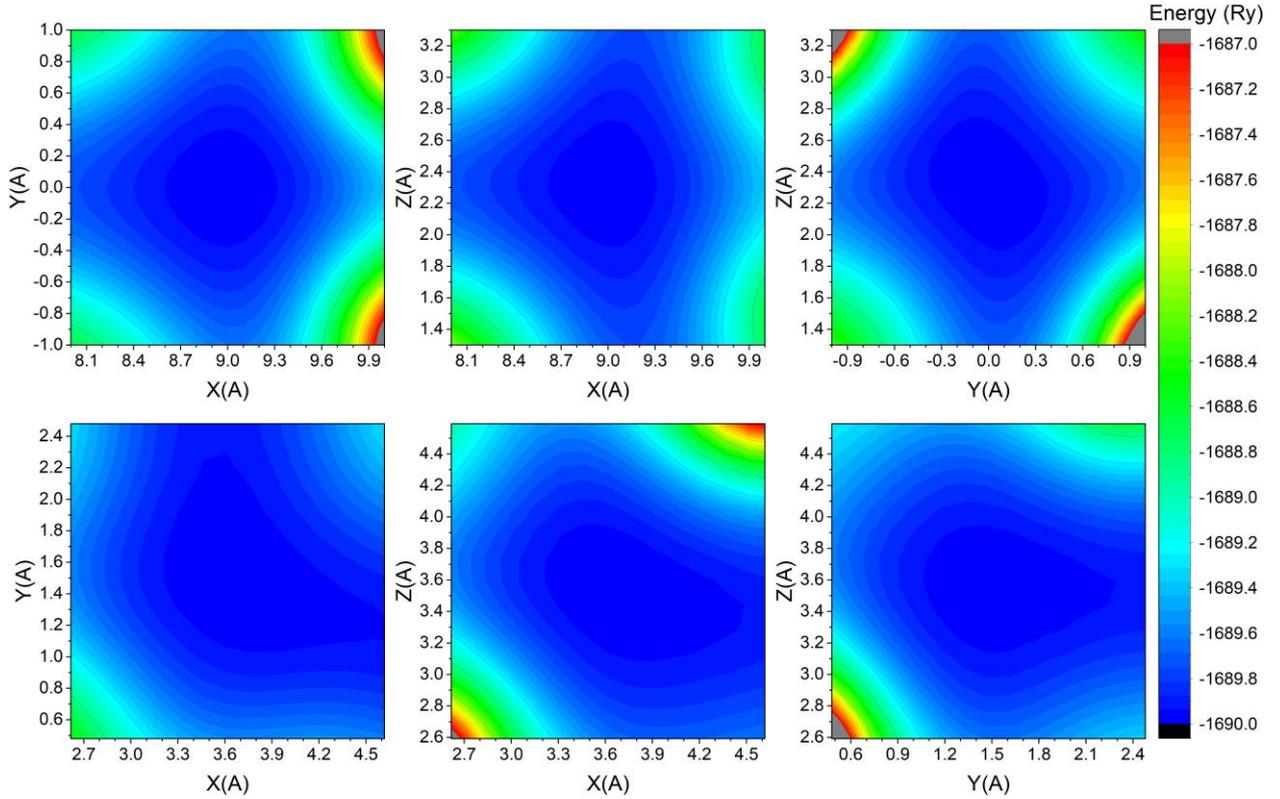

**Figure 6.** Potential energy surfaces of Ga (upper row) and O (lower row) atoms in Ia-3 phase of $Ga_2O_3$ at threshold electronic temperature at the initial time instant.

Increase of the electronic temperature to the values around the superionic threshold itself does not seem to change the potential energy surface qualitatively – it simply makes the potential well shallower (see Figure 6). It is important that an oxygen potential energy surface seems to change more drastically: one can see that oxygen energy isosurfaces, which fit inside Figure 5, went out of bounds in Figure 6 while gallium energy isosurfaces changed only slightly.

Combining obtained information, we can assume the following path of nonthermal transitions into a superionic state in the investigated compounds. The transition starts from a fast displacement of O atoms because of shallow and highly asymmetric potential well [26]. Moving oxygen atoms change the potential energy surface profile of metal atoms "dragging" them away from the equilibrium positions of an ambient crystal. This "dragging" continues until metal atoms find positions not hindering further oxygen atoms flow. After that, although the metal mean displacement remains almost constant, atoms may strongly oscillate around new positions as it is seen from mean displacements along the lattice vectors (XYZ axes).

For the observed materials exhibiting transition into a superionic state, the mean displacement saturation level at the threshold electronic temperature for metal atoms is usually around ~0.6 Å while oxygen atoms are displaced for >0.8 Å at 500 fs after the electronic temperature increase. This saturation of the metallic atoms displacements may result from energy levels changes during the transition: when a sufficient amount of metal levels turns its behavior from antibonding to bonding, metal atoms slow down settling in a new sublattice.

Summarizing the data obtained, we can suggest an algorithm helping to predict a possibility of nonthermal transition into superionic states for a material of interest.

First, the composition of bonding and antibonding energy levels in the considered material should be analyzed. If the top of the valence band and the bottom of the conduction band consist from, correspondingly, bonding and antibonding energy levels of one atomic species, this means that during fs-laser excitation electrons will leave bonding and occupy antibonding levels of these atoms shattering the corresponding sublattice. The energy levels of another atomic species will be mostly untouched. Mixed energy levels compositions most probably will lead to destabilization of both sublattices leading to nonthermal melting or solid-solid phase transition.

Second, material properties may indicate a presence of shallow or asymmetric potential well for both atomic species (instead of only one of them), which would preclude superionic state formation. For example, a low melting point, thermal instability or strong anisotropy in thermal conductivity may indicate such an asymmetry of the potential. A calculation of potential energy surfaces is the best way of such checking. An asymmetric or shallow potential for metallic subsystem may induce an additional instability upon excitations for a sublattice that otherwise would be stable, and may break the conditions favorable for the transitions into a superionic state.

We note, however, that this algorithm should be considered as a preliminary one. Its improvement is a task for future studies and an experimental validation.

We presume that non-oxide III-VI group materials should be capable of transforming into a superionic state. However, the majority of these compounds have defective structures [27] or even controversial information about atomic positions [28]. This makes them challenging for first-principles calculations [29] and requires a separate dedicated study.

## 5. Conclusion

We established threshold electronic temperatures triggering nonthermal phase transitions and their types (solid-liquid or solid-superionic) in a number of group III-metal oxides. We demonstrated that the majority of considered materials exhibit nonthermal transitions into a superionic phase where oxygen exhibits a liquid-like behavior in contrast to the metallic sublattice remaining in a solid state.

We analyzed electronic structures of these materials and concluded that a unique combination of bonding and antibonding states may be responsible for such a behavior, which can be used as an indicator of a possibility of a transition into a superionic state. The behavior of two materials (C2/m phase of $Ga_2O_3$ and R-3c phase of $In_2S_3$) under high electronic excitation showed that the initial asymmetries of the potential energy surfaces of metallic and oxygen atoms hindered their transition into a superionic state.

### Acknowledgements

This work has been carried out using computing resources of the Federal collective usage center Complex for Simulation and Data Processing for Mega-science Facilities at NRC "Kurchatov Institute", http://ckp.nrcki.ru/, as well as computing resources of GSI Helmholtzzentrum (Darmstadt, Germany).

Partial support from project No. 16 APPA (GSI) funded by the Ministry of Science and Higher Education of the Russian Federation, is acknowledged. NM gratefully acknowledges financial support from the Czech Ministry of Education, Youth and Sports (Grants No. LTT17015 and LM2018114). This work benefited from networking activities carried out within the EU funded COST Action CA17126 (TUMIEE) and represents a contribution to it.

**Supplementary materials for**

**"Superionic states formation in group III oxides irradiated with ultrafast lasers"**

R.A. Voronkov[1*], N. Medvedev[2,3], A.E. Volkov[1]

[1]*P. N. Lebedev Physical Institute of the Russian Academy of Sciences, Leninskij pr., 53,119991 Moscow, Russia;*

[2]*Institute of Physics, Czech Academy of Sciences, Na Slovance 2, 182 21 Prague 8, Czech Republic;*

[3]*Institute of Plasma Physics, Czech Academy of Sciences, Za Slovankou 3, 182 00 Prague 8, Czech Republic;*

## I.    Information about nonthermal transitions

In this file, detailed information about nonthermal transitions from the main text is presented. In order to justify thresholds and transition type claimed in the manuscript, for each material mean atomic displacements and calculated X-ray diffraction (XRD) patterns are shown in Section I for the electronic temperatures below the transition threshold as well as those for the threshold. XRD patterns of the simulated supercell are calculated via VESTA software. Mean displacements along lattice vectors at the threshold temperature are also presented to provide additional details about atomic kinetics during transitions. Projected electronic densities of states for each material are shown in Section II.

# 1. Al₂O₃ Ia-3

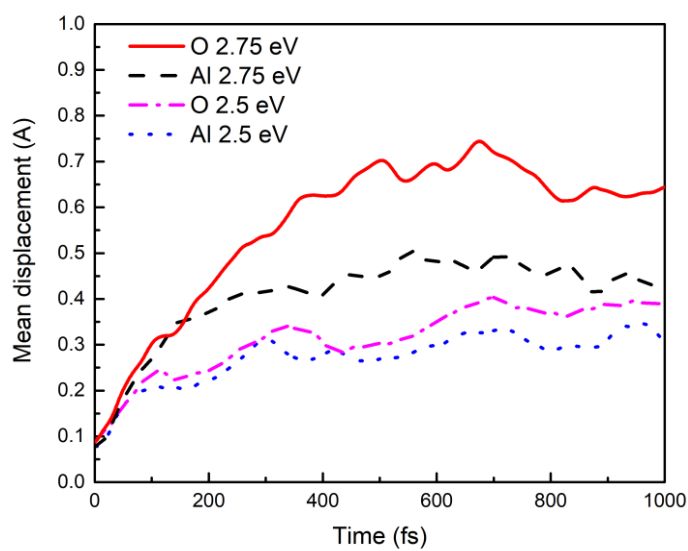

**Figure 1.1.1.** Mean atomic displacements at the electronic temperature (2.5 eV) below the threshold and at the threshold electronic temperature (2.75 eV).

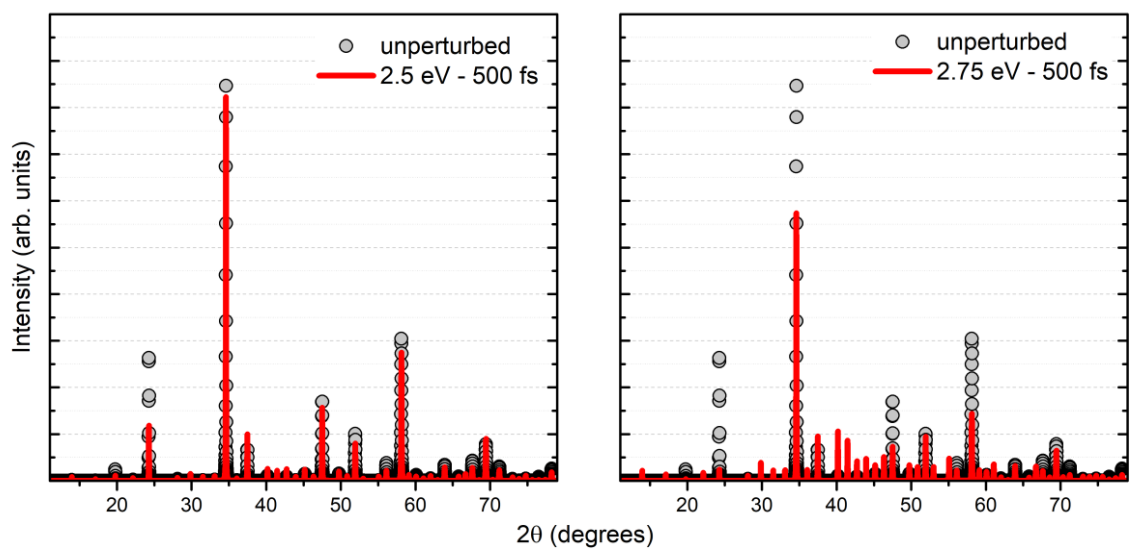

**Figure 1.1.2.** XRD patterns (λ=1.5406 Å) of the simulated supercell at $T_e$ = 2.5 eV (below threshold) and $T_e$ = 2.75 eV (threshold) at the initial and final time instants.

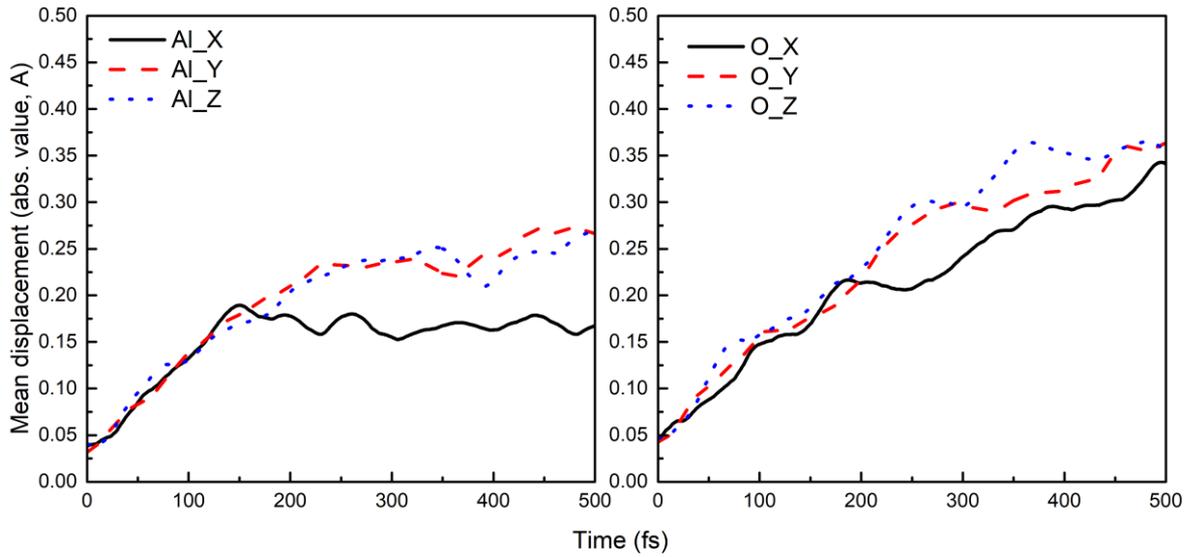

**Figure 1.1.3.** Absolute values of mean atomic displacements along lattice vectors at the threshold temperature $T_e = 2.75$ eV.

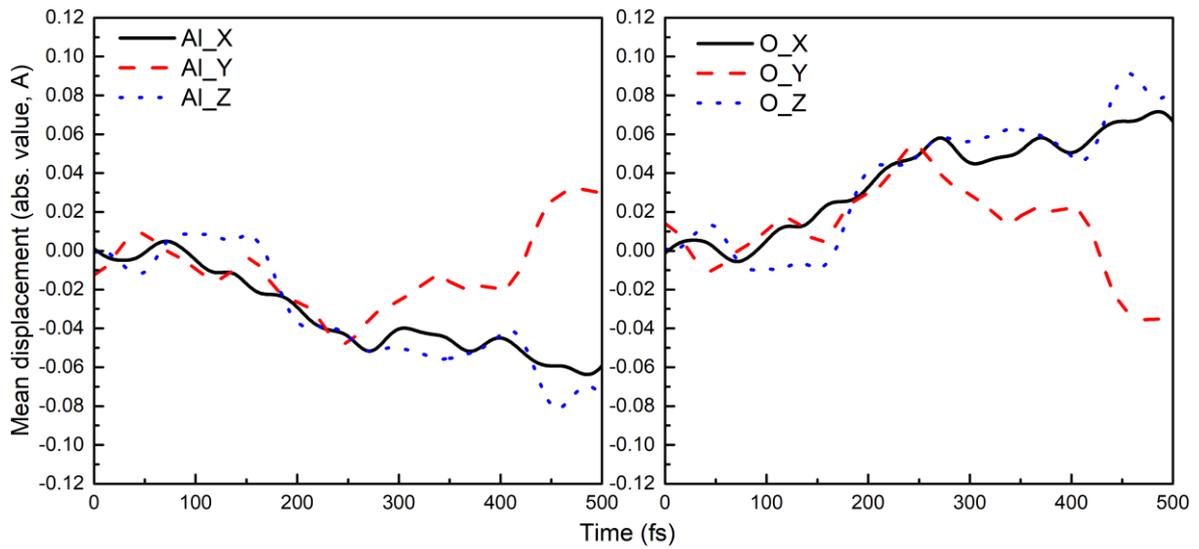

**Figure 1.1.4.** Mean atomic displacements along lattice vectors at the threshold temperature $T_e = 2.75$ eV. Negative values correspond to the movement against lattice vector direction.

## 2. Ga₂O₃ R-3c

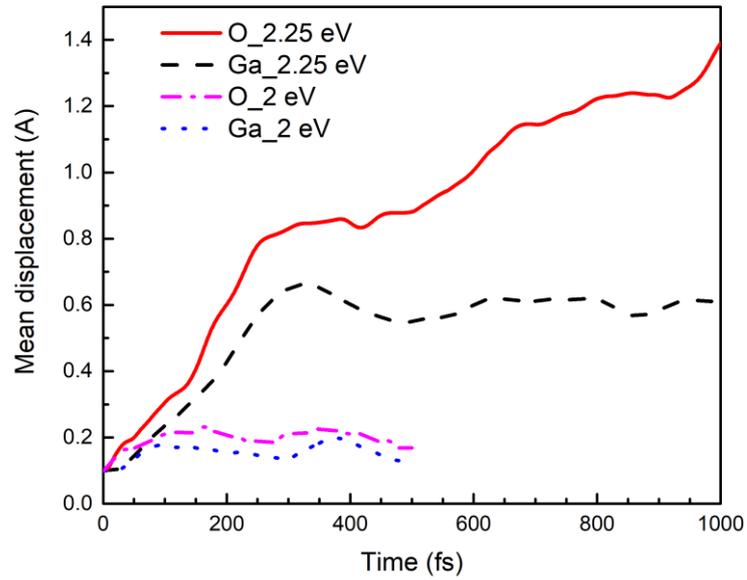

**Figure 1.2.1.** Mean atomic displacements at the electronic temperature (2 eV) below the threshold and at the threshold electronic temperature (2.25 eV).

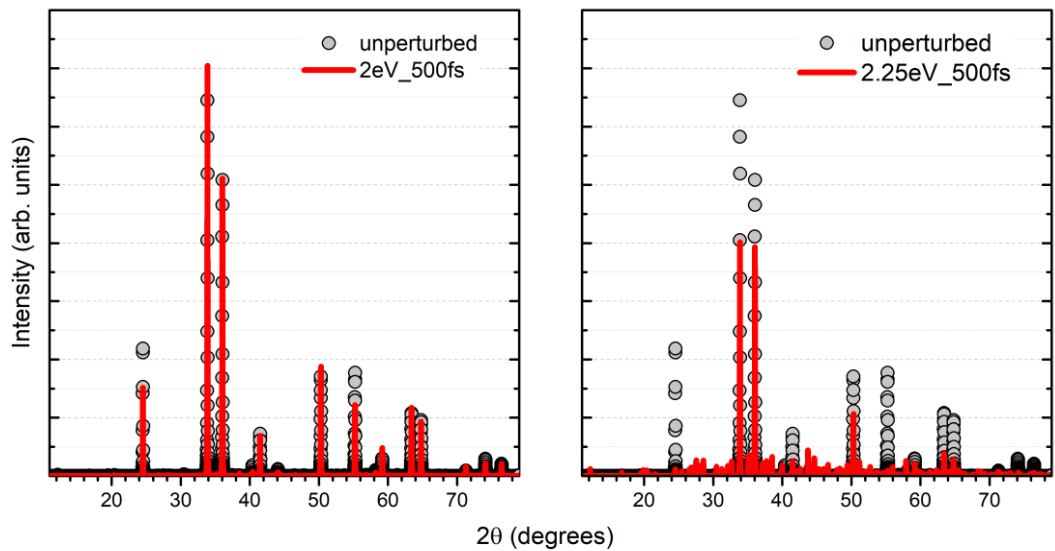

**Figure 1.2.2.** XRD patterns (λ=1.5406 Å) of the simulated supercell at $T_e$ = 2 eV (below threshold) and $T_e$ = 2.25 eV (threshold) at the initial and final time instants.

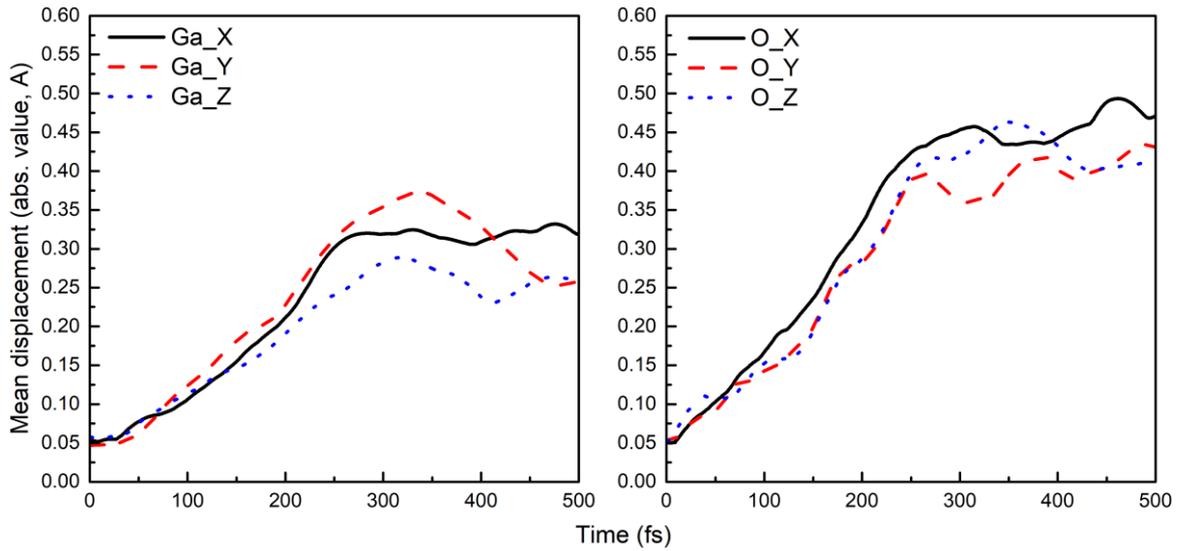

**Figure 1.2.3.** Absolute values of mean atomic displacements along lattice vectors at the threshold temperature $T_e = 2.25$ eV.

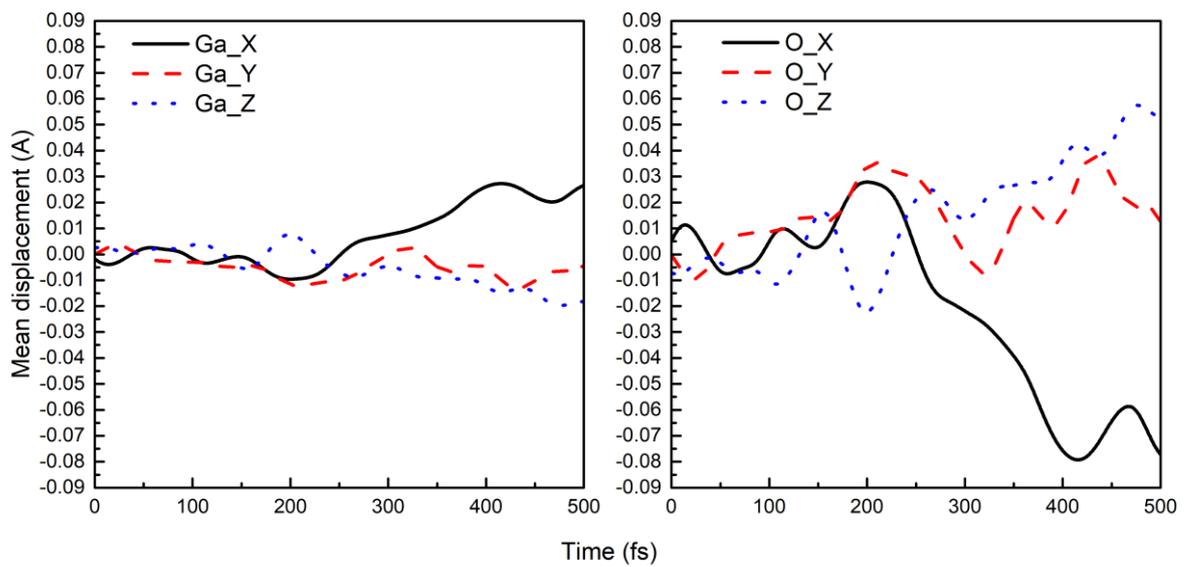

**Figure 1.2.4.** Mean atomic displacements along lattice vectors at the threshold temperature $T_e = 2.25$ eV. Negative values correspond to the movement against lattice vector direction.

# 3. Ga₂O₃ Ia-3

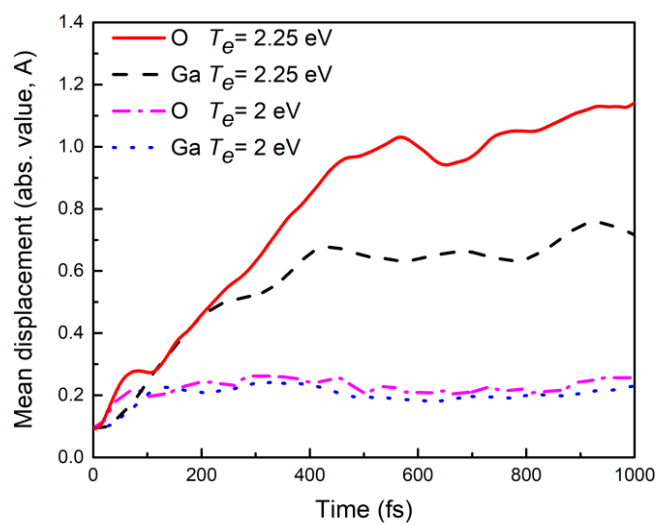

**Figure 1.3.1.** Mean atomic displacements at the electronic temperature (2 eV) below the threshold and at the threshold electronic temperature (2.25 eV).

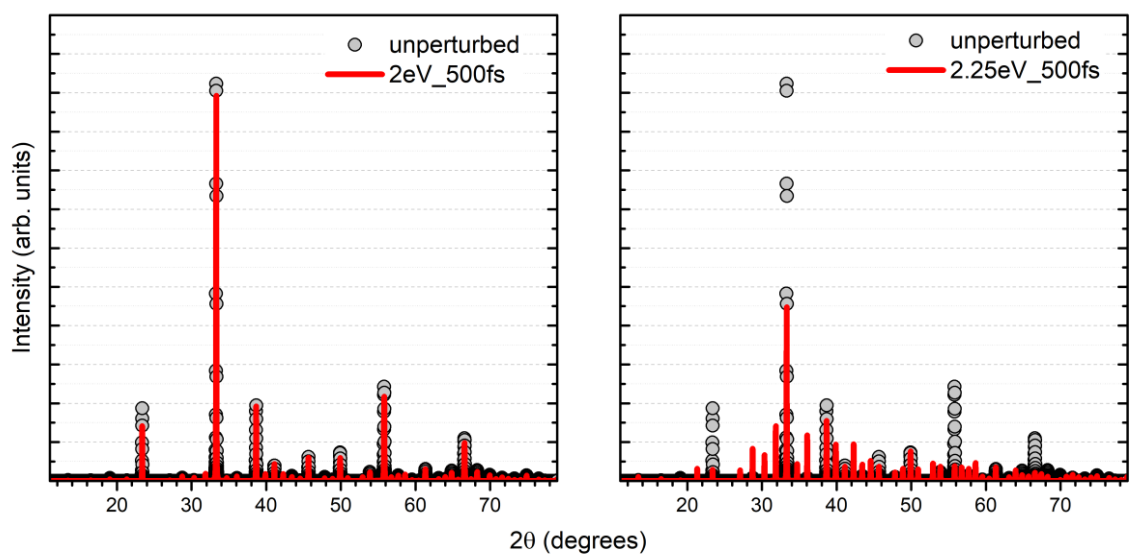

**Figure 1.3.2.** XRD patterns ($\lambda$=1.5406 Å) of the simulated supercell at $T_e$ = 2 eV (below threshold) and $T_e$ = 2.25 eV (threshold) at the initial and final time instants.

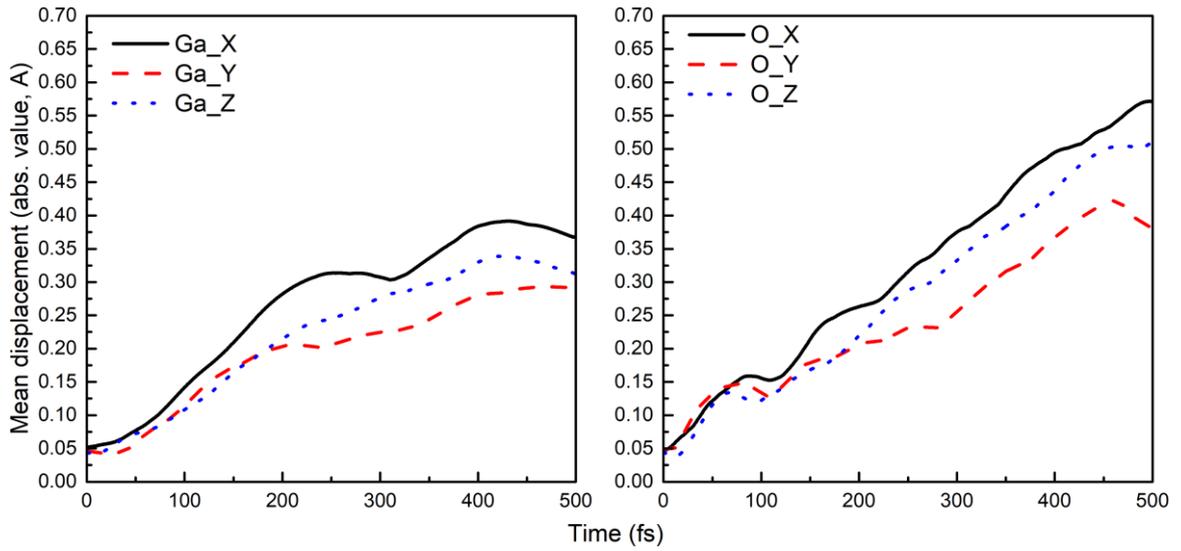

**Figure 1.3.3.** Absolute values of mean atomic displacements along lattice vectors at the threshold temperature $T_e = 2.25$ eV.

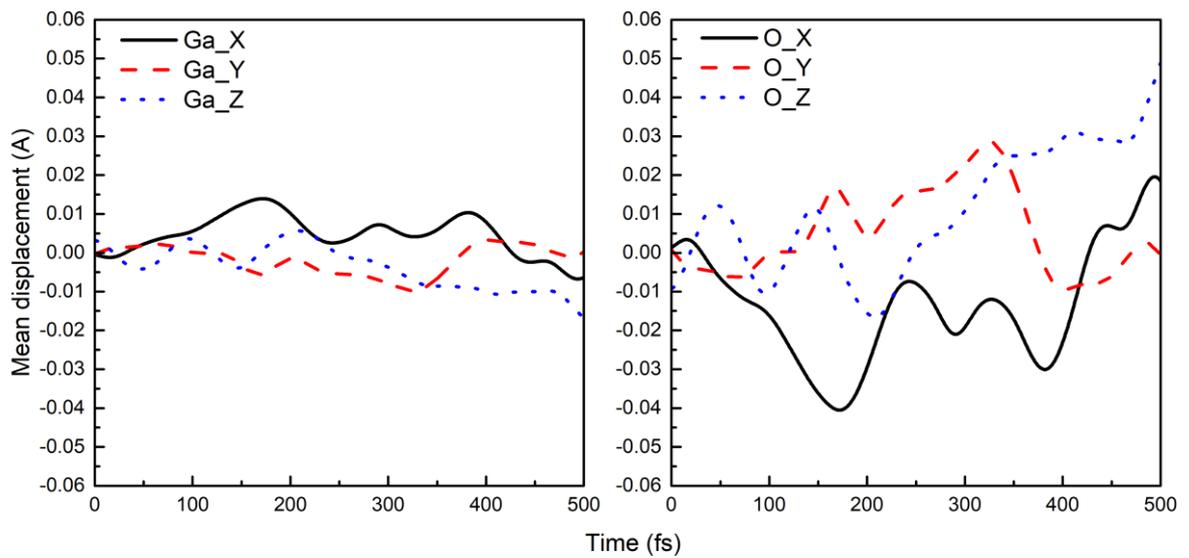

**Figure 1.3.4.** Mean atomic displacements along lattice vectors at the threshold temperature $T_e = 2.25$ eV. Negative values correspond to the movement against lattice vector direction.

# 4. Ga₂O₃ C2/m

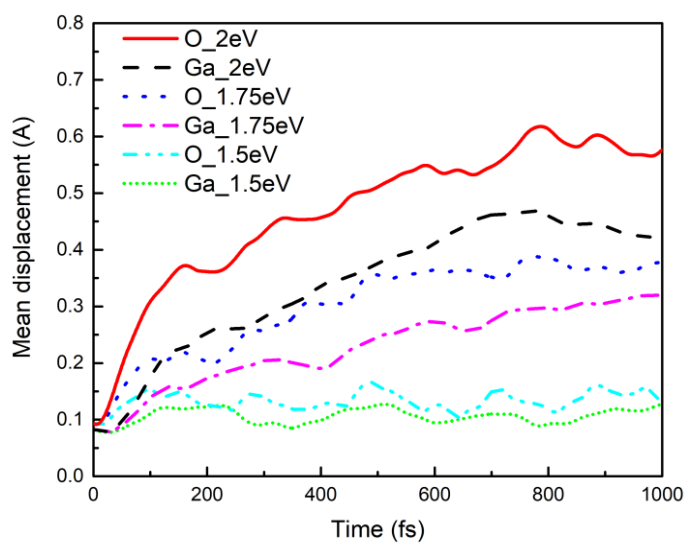

**Figure 1.4.1.** Mean atomic displacements at the electronic temperature (1.5 eV) below the threshold, at the threshold electronic temperature (1.75 eV) and the electronic temperature above threshold (2 eV).

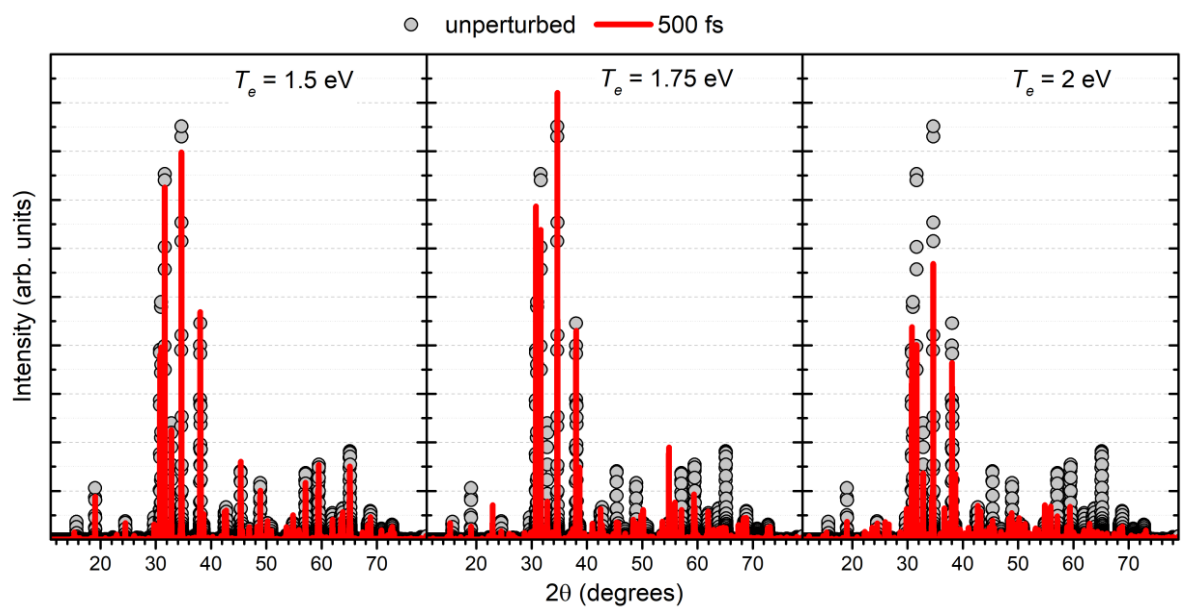

**Figure 1.4.2.** XRD patterns (λ=1.5406 Å) of the simulated supercell at the electronic temperature (1.5 eV) below the threshold, at the threshold electronic temperature (1.75 eV) and the electronic temperature above threshold (2 eV) at the initial and final time instants.

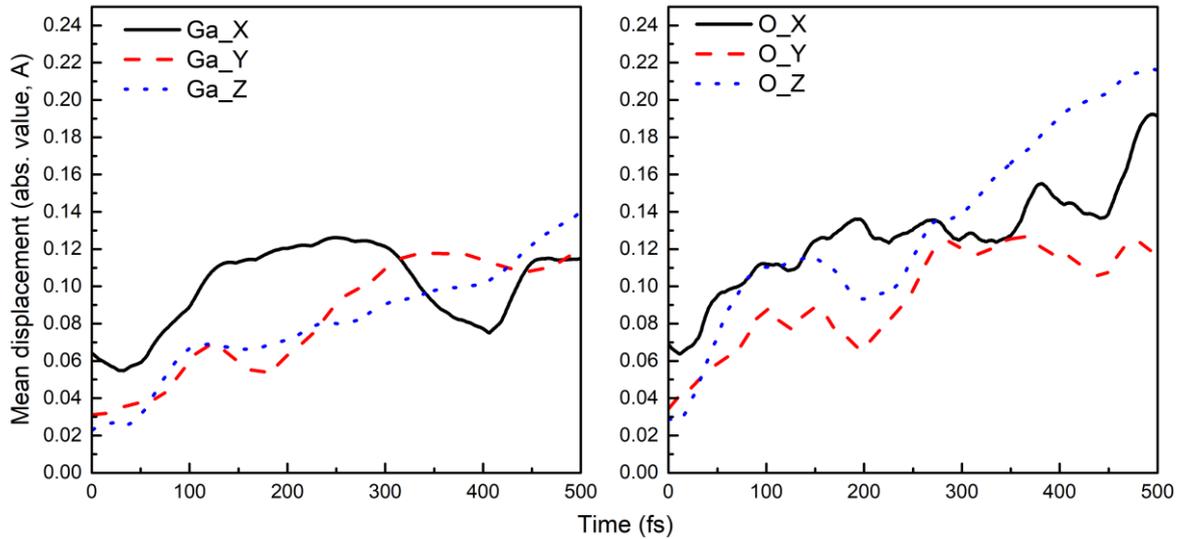

**Figure 1.4.3.** Absolute values of mean atomic displacements along lattice vectors at the threshold temperature $T_e = 1.75$ eV.

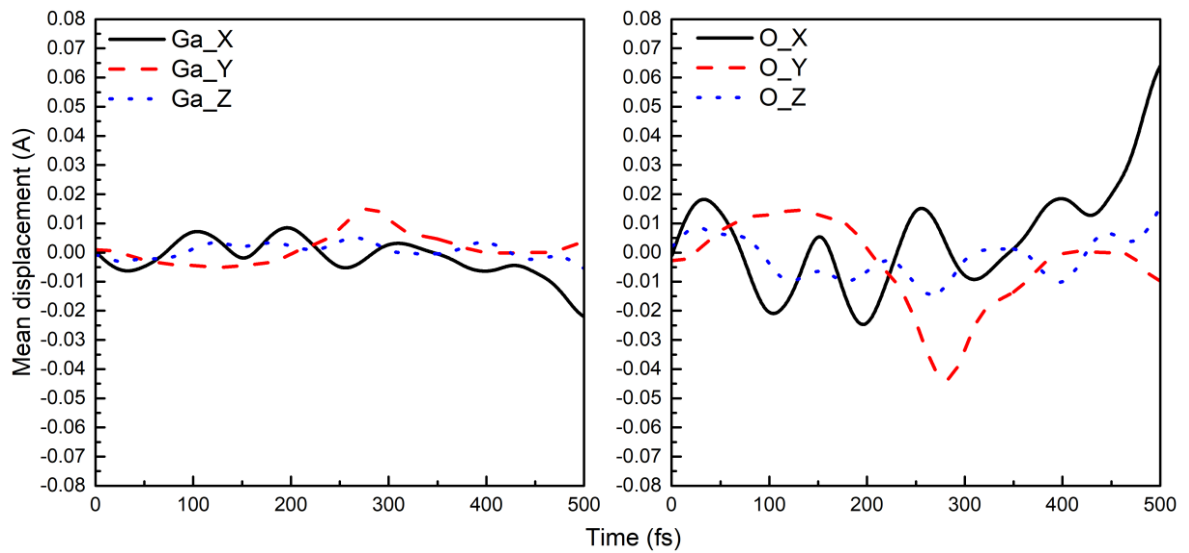

**Figure 1.4.4.** Mean atomic displacements along lattice vectors at the threshold temperature $T_e = 1.75$ eV. Negative values correspond to the movement against lattice vector direction.

## 5. In₂O₃ R-3c

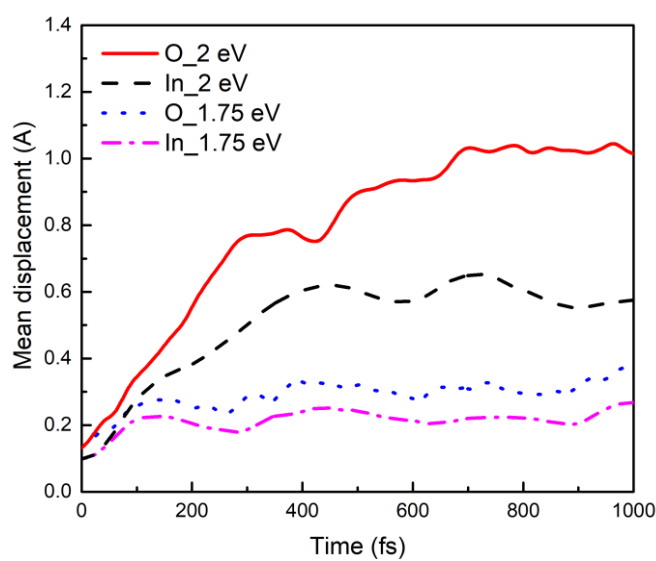

**Figure 1.5.1.** Mean atomic displacements at the electronic temperature (1.75 eV) below the threshold and at the threshold electronic temperature (2 eV).

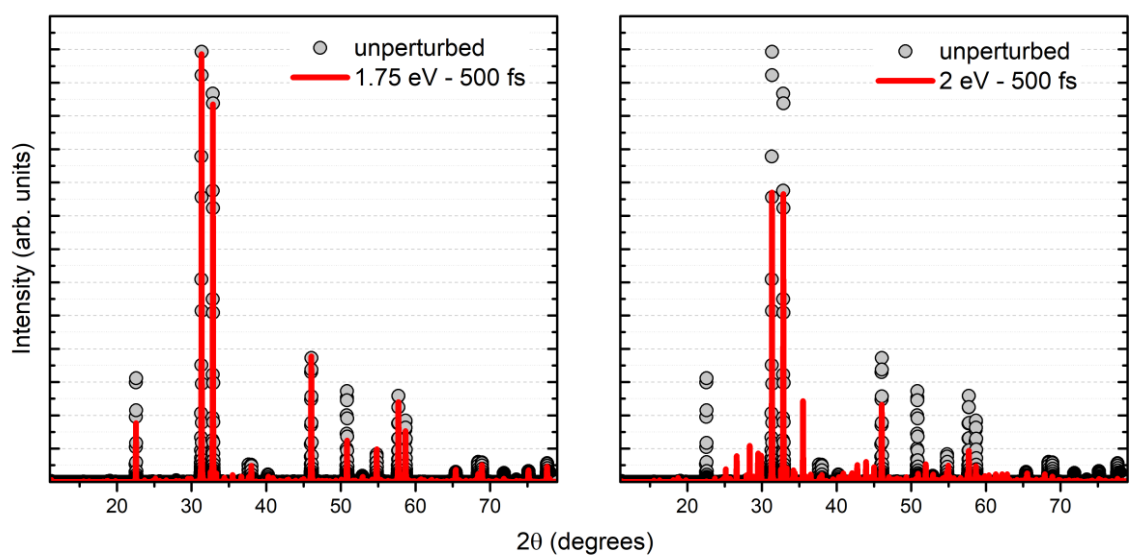

**Figure 1.5.2.** XRD patterns ($\lambda$=1.5406 Å) of the simulated supercell at $T_e$ = 1.75 eV (below threshold) and $T_e$ = 2 eV (threshold) at the initial and final time instants.

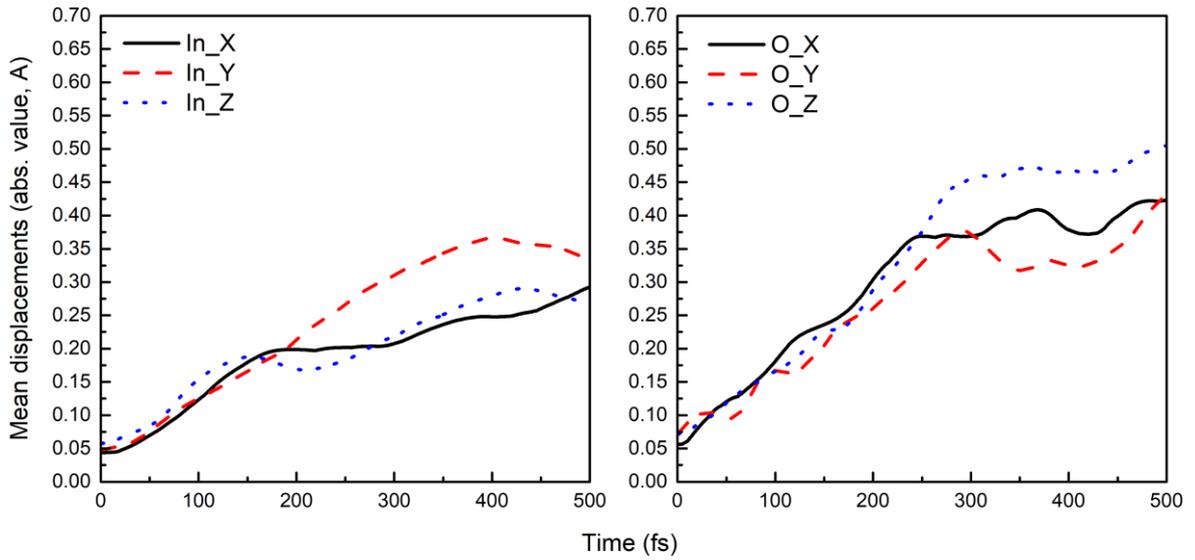

**Figure 1.5.3.** Absolute values of mean atomic displacements along lattice vectors at the threshold temperature $T_e = 2$ eV.

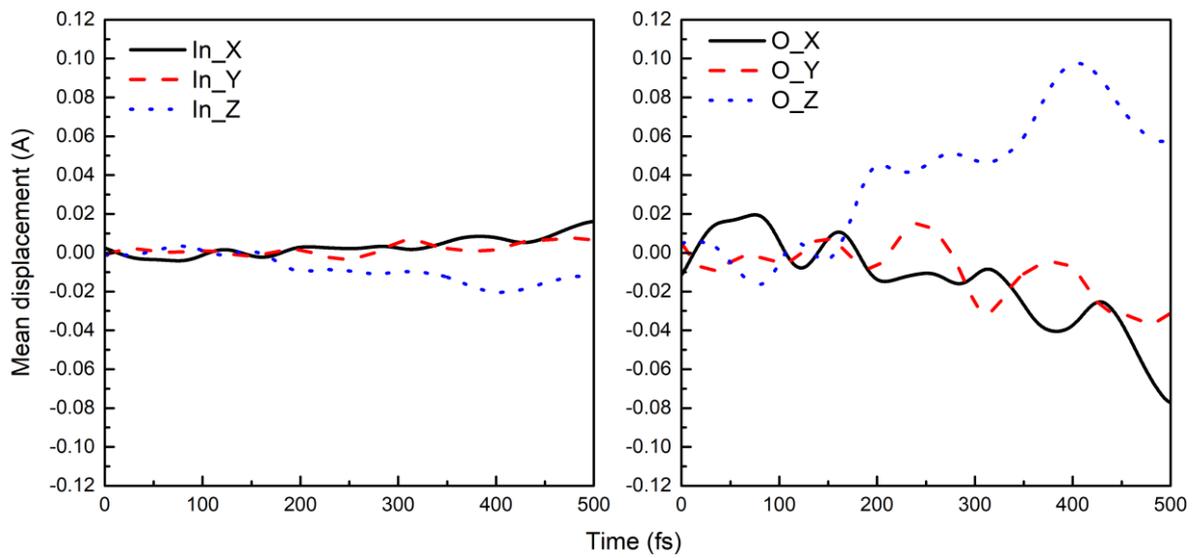

**Figure 1.5.4.** Mean atomic displacements along lattice vectors at the threshold temperature $T_e = 2$ eV. Negative values correspond to the movement against lattice vector direction.

# 6. In₂O₃ Ia-3

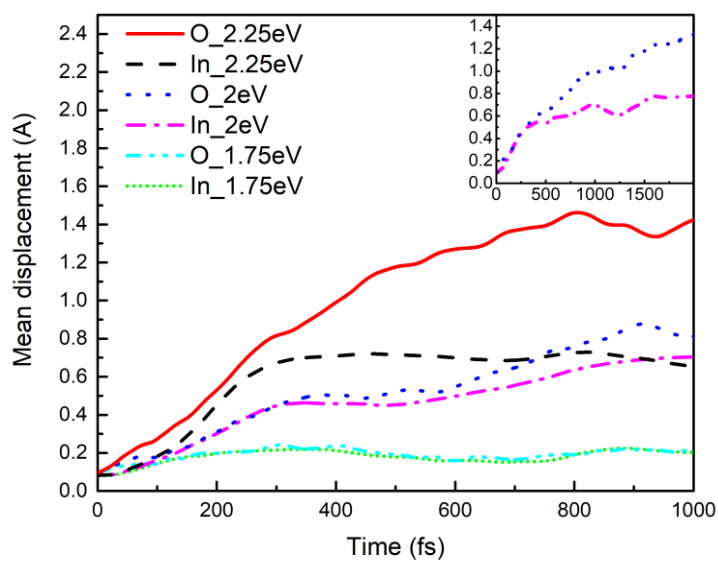

**Figure 1.6.1.** Mean atomic displacements at the electronic temperature (1.75 eV) below the threshold, at the threshold electronic temperature (2 eV) and the electronic temperature above the threshold one (2.25 eV).

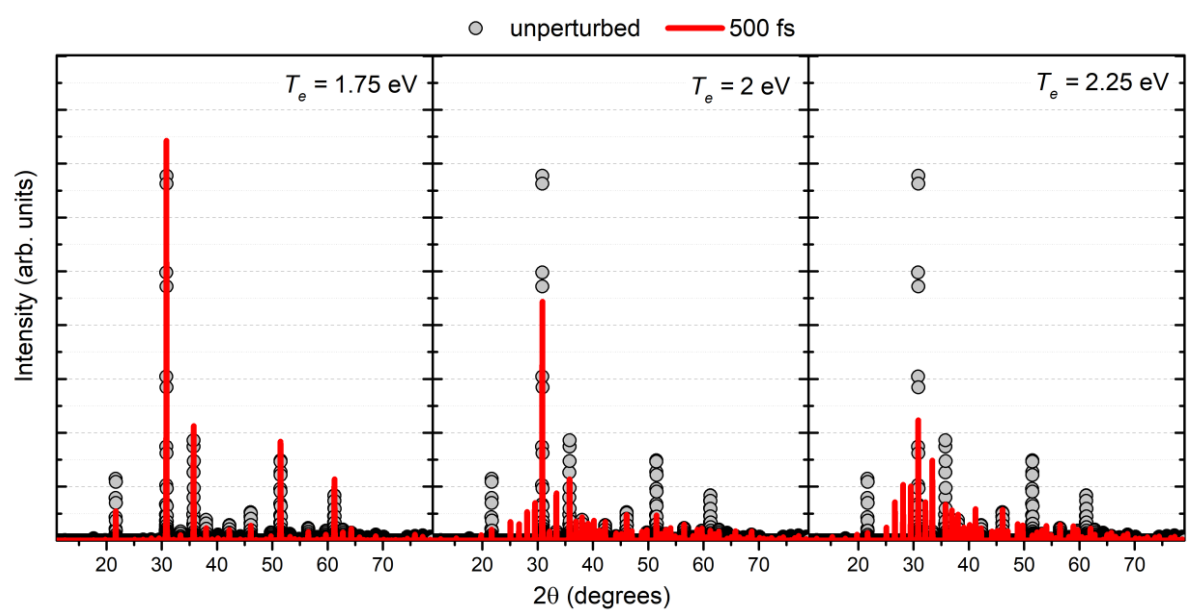

**Figure 1.6.2.** XRD patterns (λ=1.5406 Å) of the simulated supercell at the electronic temperature (1.75 eV) below the threshold, at the threshold electronic temperature (2 eV) and the electronic temperature above threshold (2.25 eV) at the initial and final time instants.

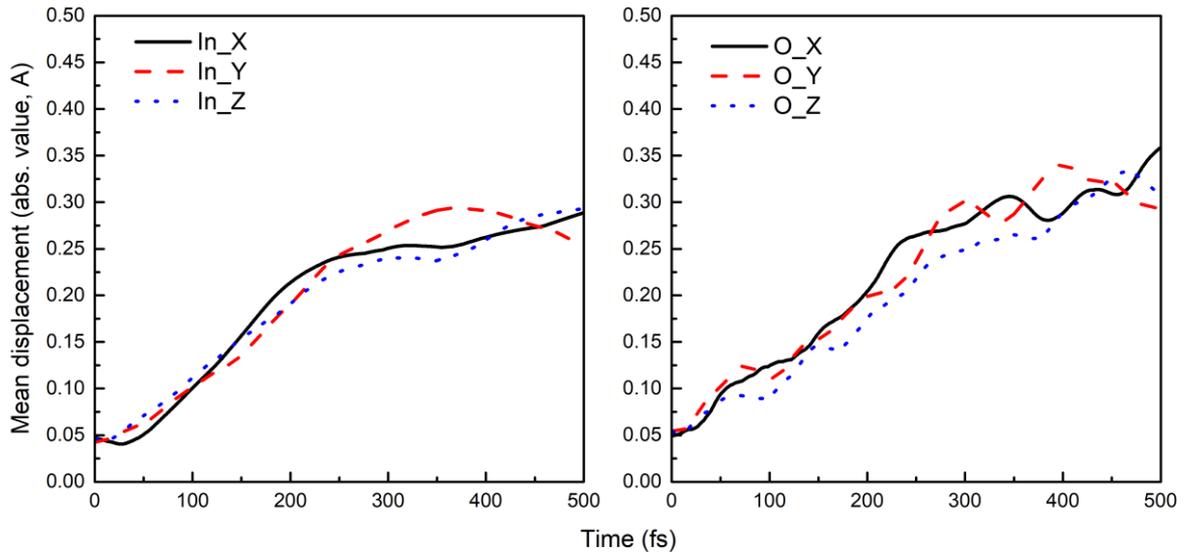

**Figure 1.6.3.** Absolute values of mean atomic displacements along lattice vectors at the threshold temperature $T_e = 2$ eV.

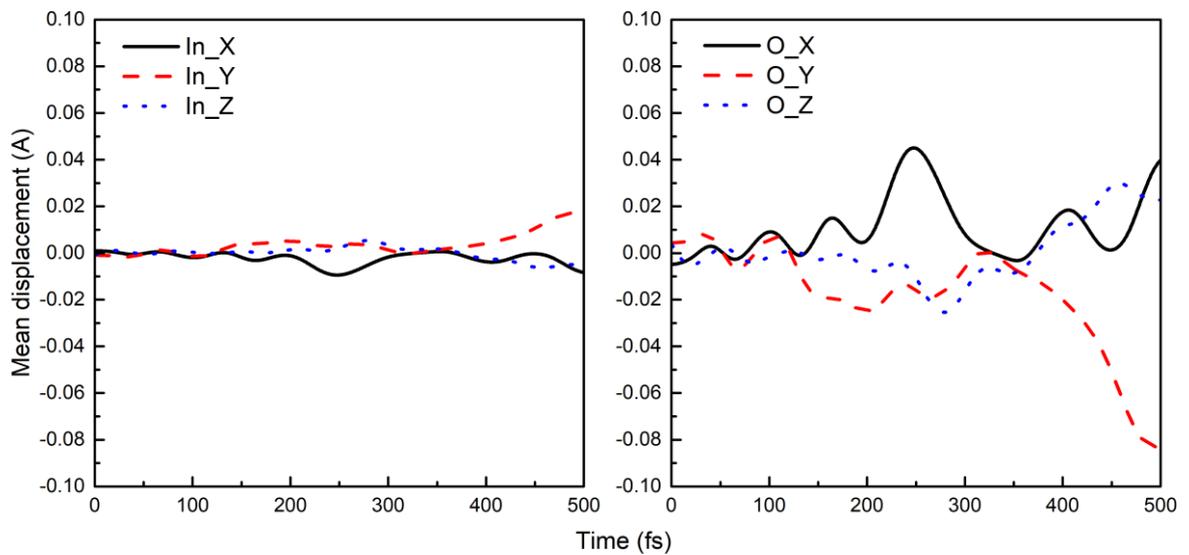

**Figure 1.6.4.** Mean atomic displacements along lattice vectors at the threshold temperature. Negative values correspond to the movement against lattice vector direction.

## 7. In₂S₃ R-3c

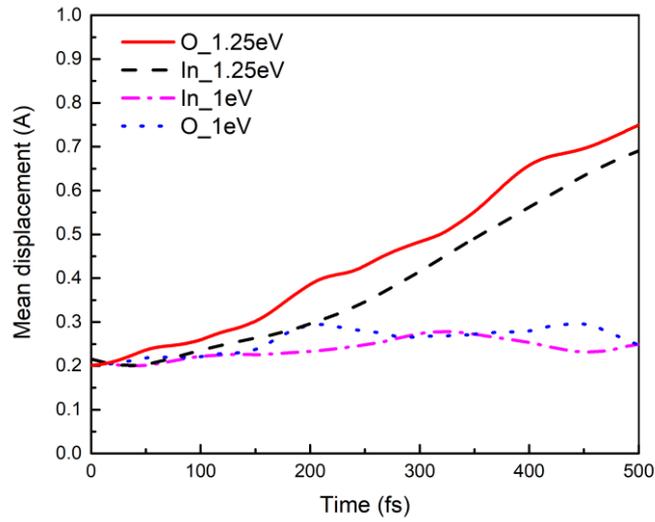

**Figure 1.7.1.** Mean atomic displacements at the electronic temperature (1 eV) below the threshold and at the threshold electronic temperature (1.25 eV).

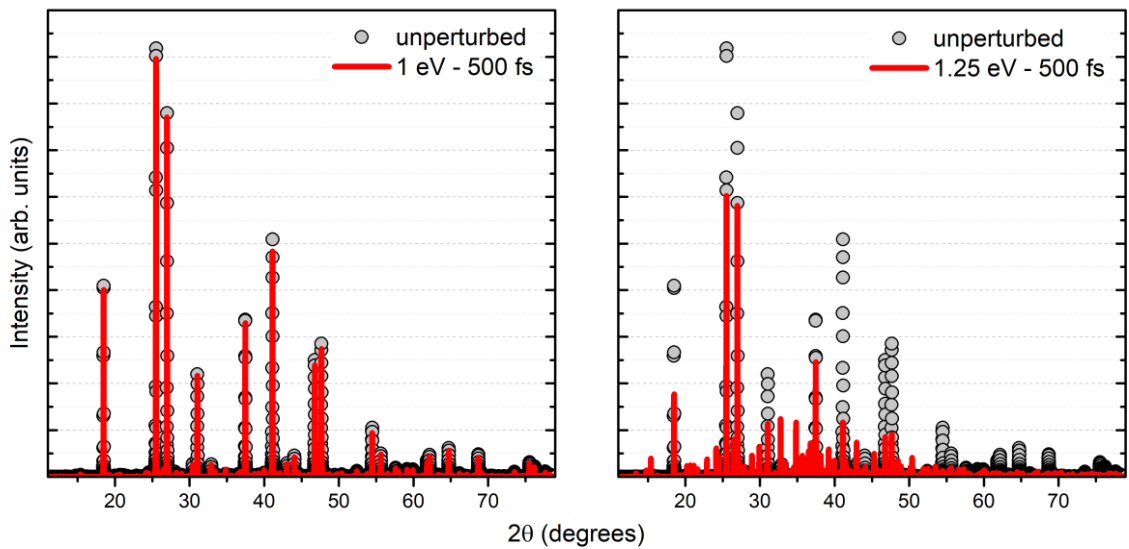

**Figure 1.7.2.** XRD patterns ($\lambda$=1.5406 Å) of the simulated supercell at $T_e$ = 1 eV (below threshold) and $T_e$ = 1.25 eV (threshold) at the initial and final time instants.

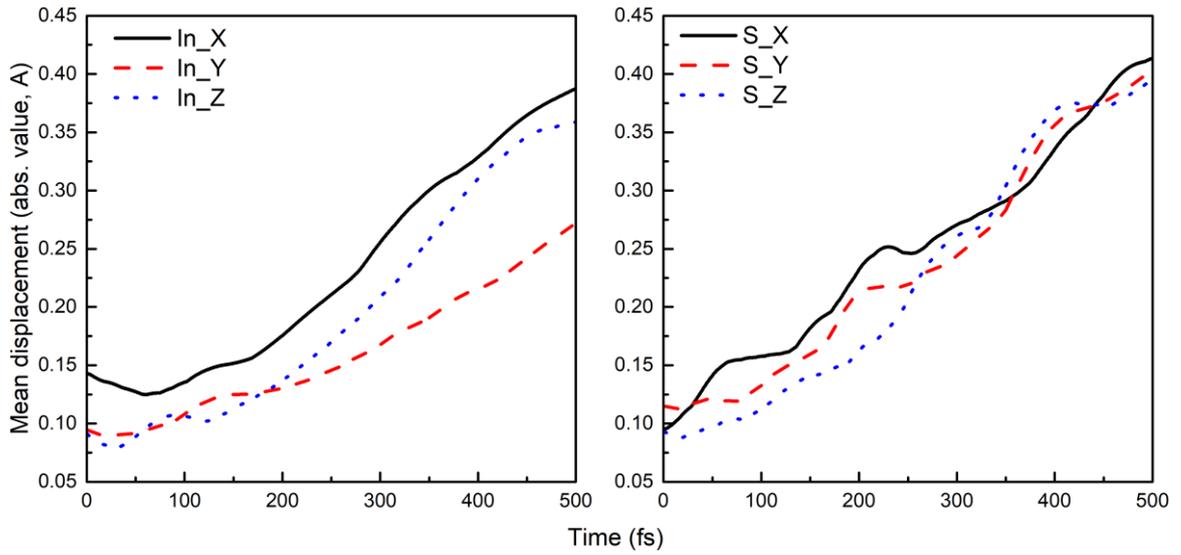

**Figure 1.7.3.** Absolute values of mean atomic displacements along lattice vectors at the threshold temperature $T_e = 1.25$ eV.

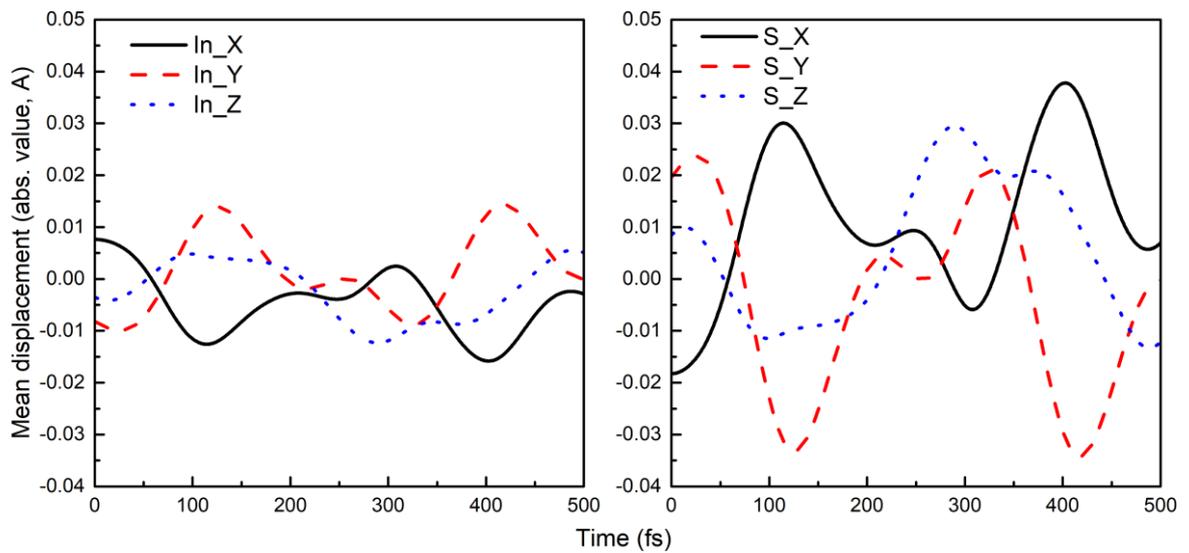

**Figure 1.7.4.** Mean atomic displacements along lattice vectors at the threshold temperature $T_e = 1.25$ eV. Negative values correspond to the movement against lattice vector direction.

# II.    Projected electronic density of states

## 1.  Total contribution of O and Me atoms to electronic DOS

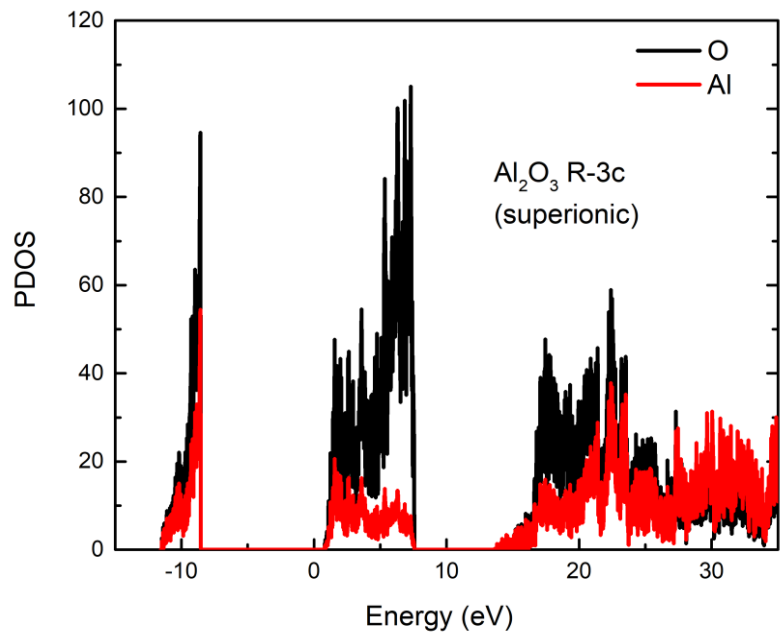

**Figure 2.1.1.** Total contribution of O and Al atoms to electronic DOS of R-3c phase of $Al_2O_3$.

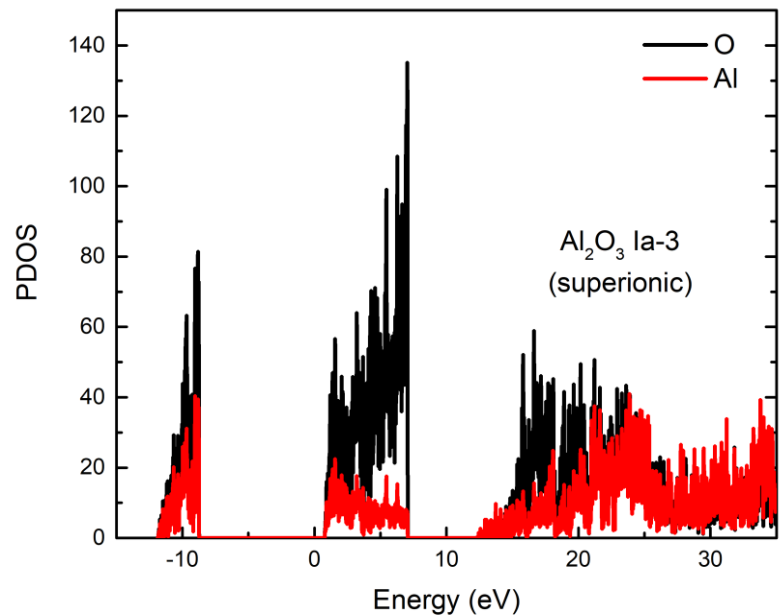

**Figure 2.1.2.** Total contribution of O and Al atoms to electronic DOS of Ia-3 phase of $Al_2O_3$.

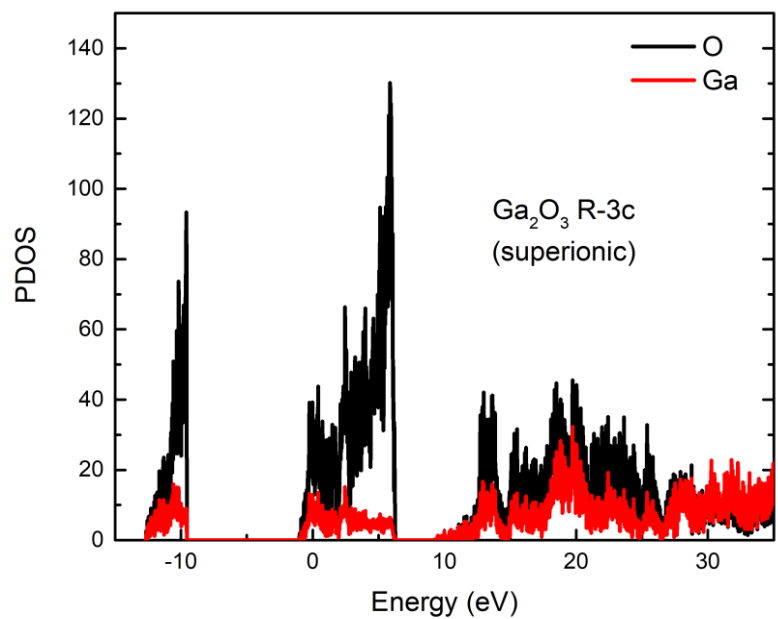

**Figure 2.1.3.** Total contribution of O and Ga atoms to electronic DOS of R-3c phase of $Ga_2O_3$.

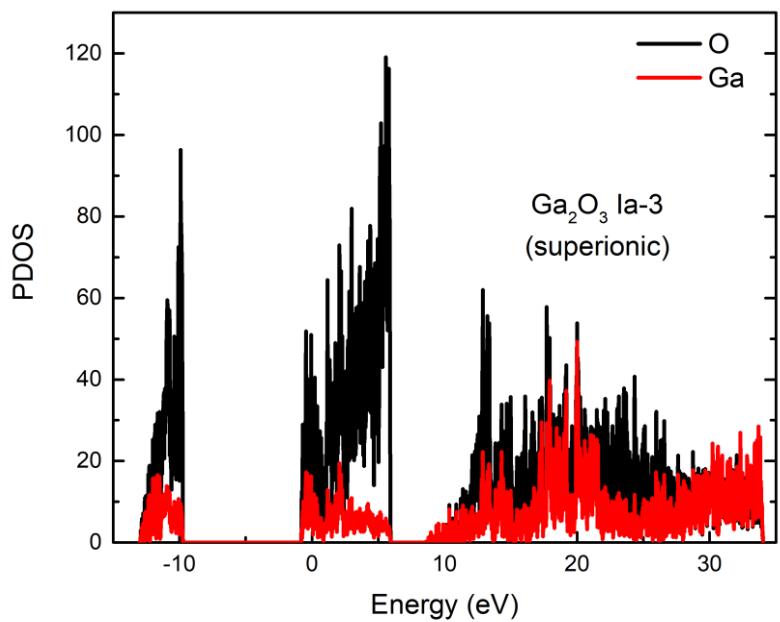

**Figure 2.1.4.** Total contribution of O and Ga atoms to electronic DOS of Ia-3 phase of $Ga_2O_3$.

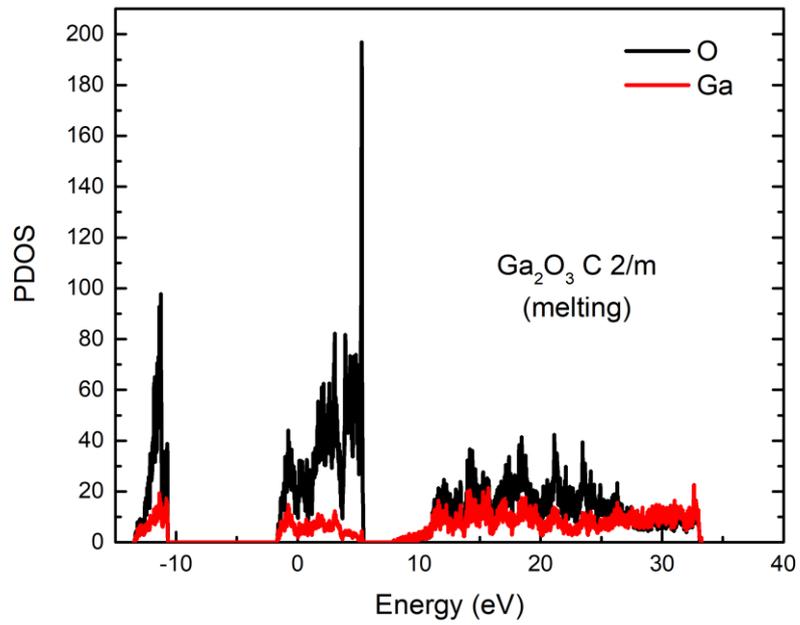

**Figure 2.1.5.** Total contribution of O and Ga atoms to electronic DOS of C2/m phase of $Ga_2O_3$.

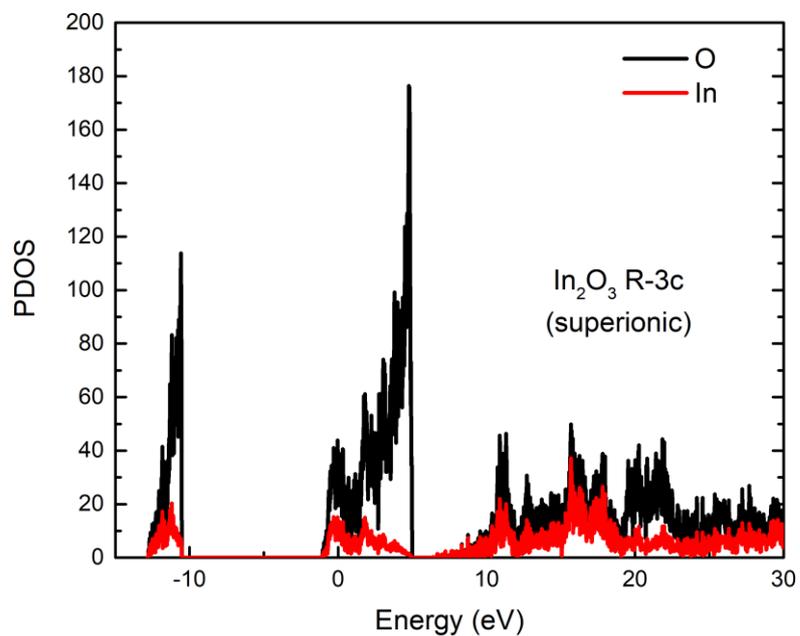

**Figure 2.1.6.** Total contribution of O and In atoms to electronic DOS of R-3c phase of $In_2O_3$.

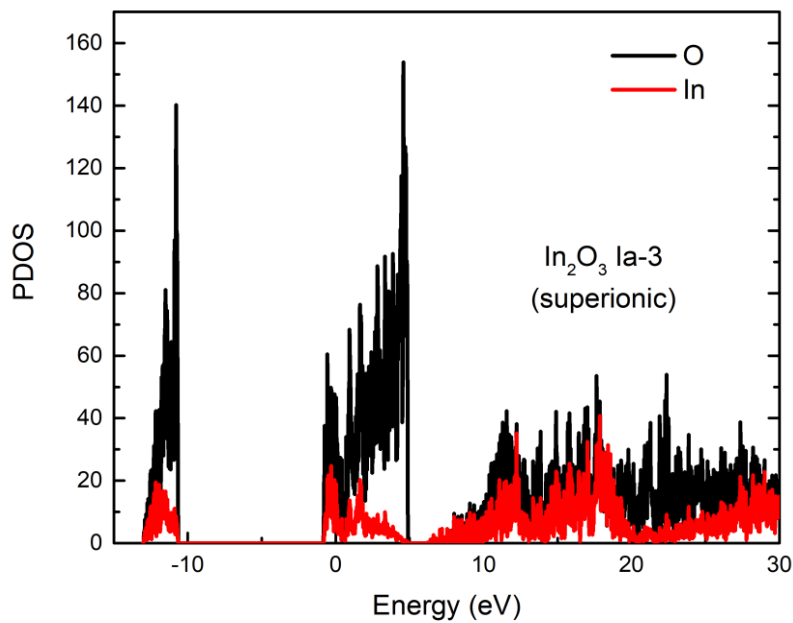

**Figure 2.1.7.** Total contribution of O and In atoms to electronic DOS of Ia-3 phase of $In_2O_3$.

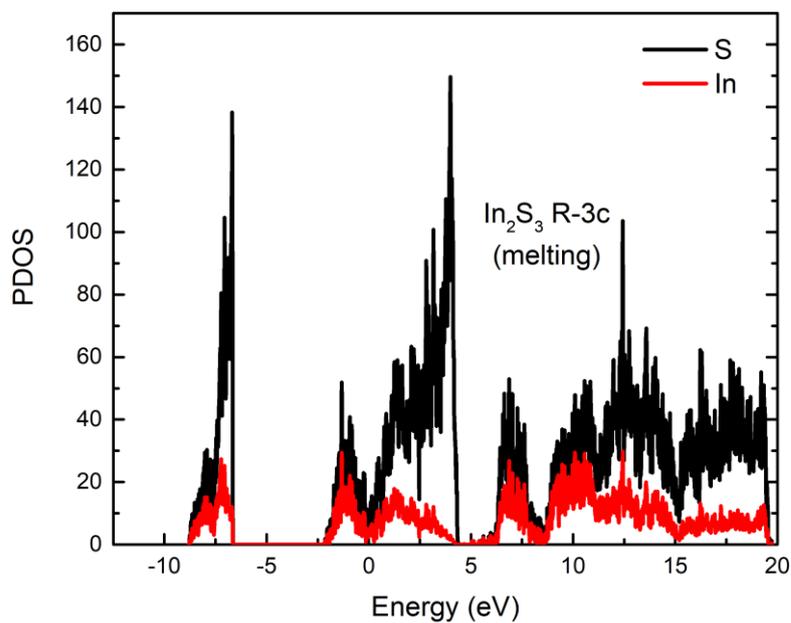

**Figure 2.1.8.** Total contribution of S and In atoms to electronic DOS of R-3c phase of $In_2S_3$.

## 2. Contribution of O orbitals to electronic DOS

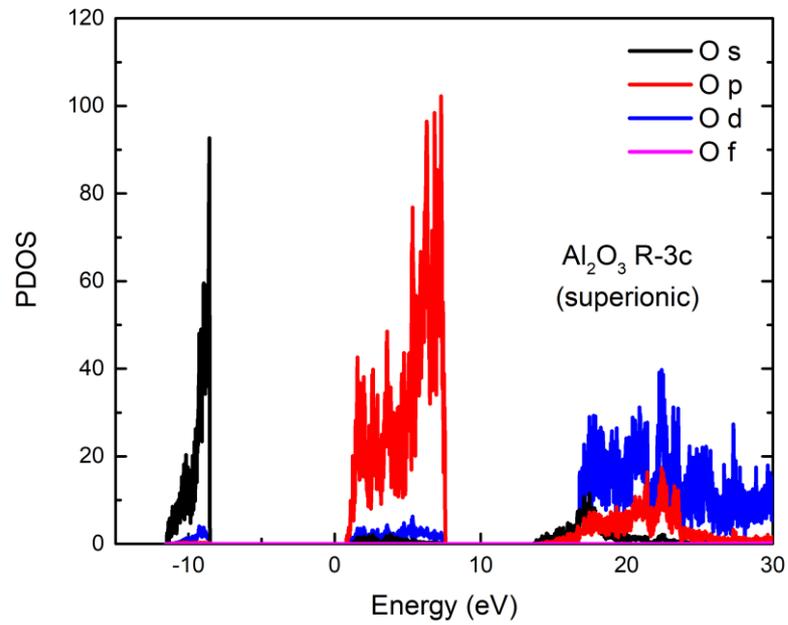

**Figure 2.2.1.** Contribution of O orbitals to electronic DOS of R-3c phase of $Al_2O_3$.

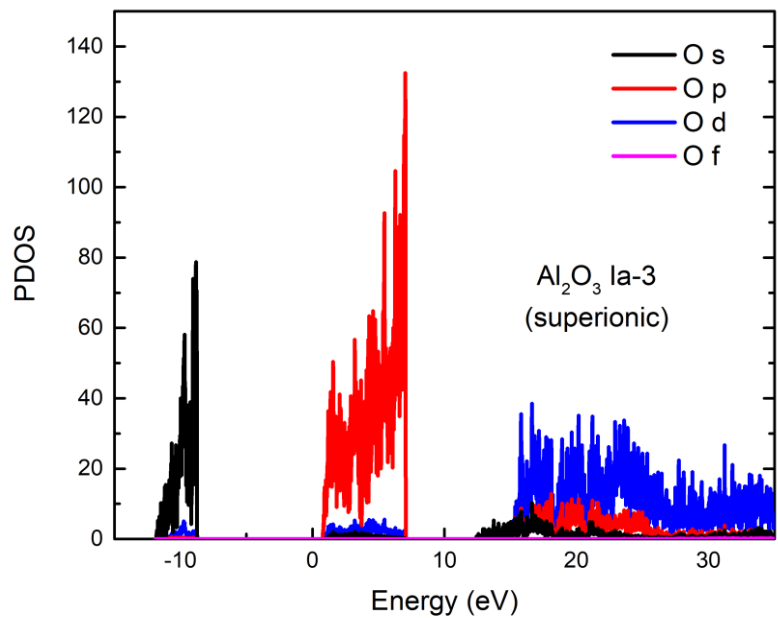

**Figure 2.2.2.** Contribution of O orbitals to electronic DOS of Ia-3 phase of $Al_2O_3$.

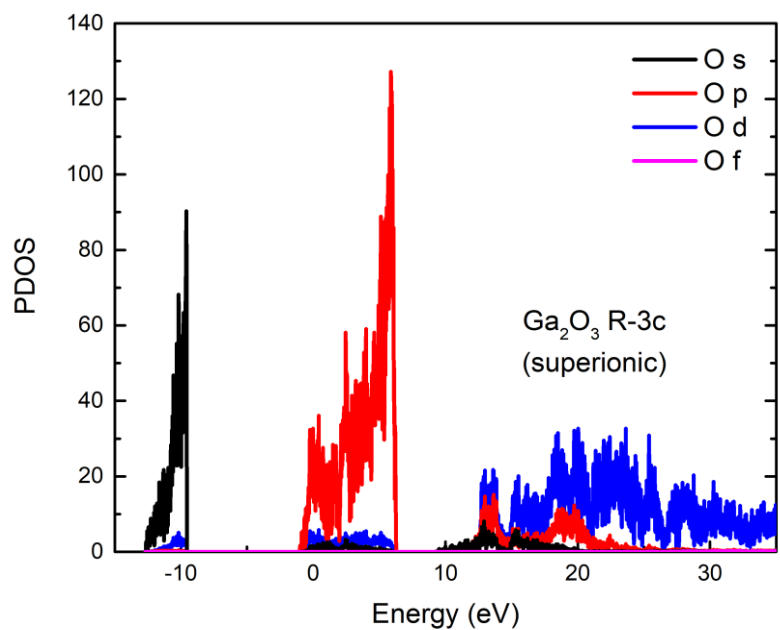

**Figure 2.2.3.** Contribution of O orbitals to electronic DOS of R-3c phase of $Ga_2O_3$.

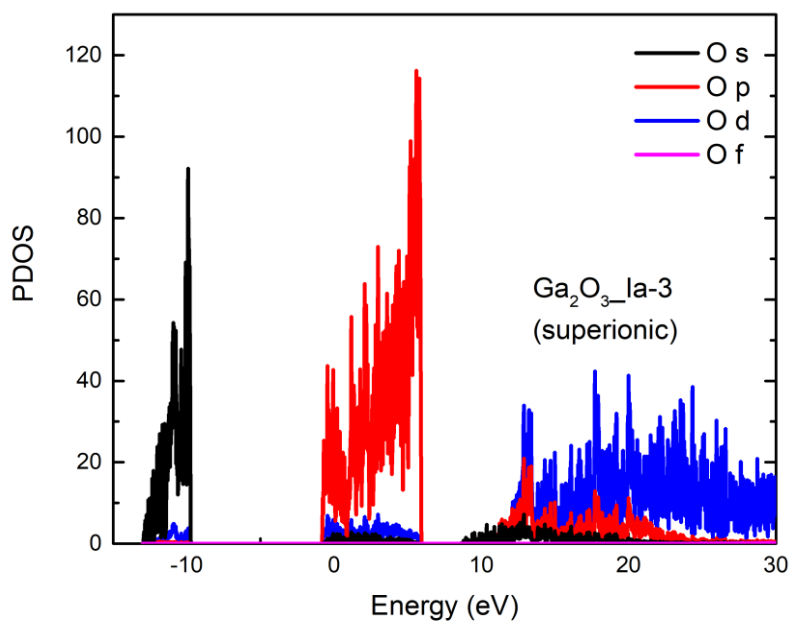

**Figure 2.2.4.** Contribution of O orbitals to electronic DOS of Ia-3 phase of $Ga_2O_3$.

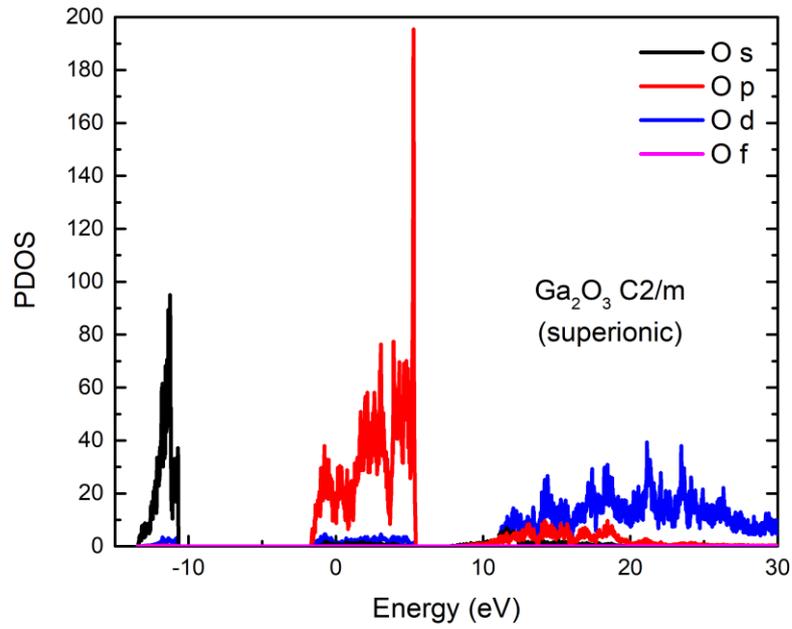

**Figure 2.2.5.** Contribution of O orbitals to electronic DOS of C2/m phase of Ga$_2$O$_3$.

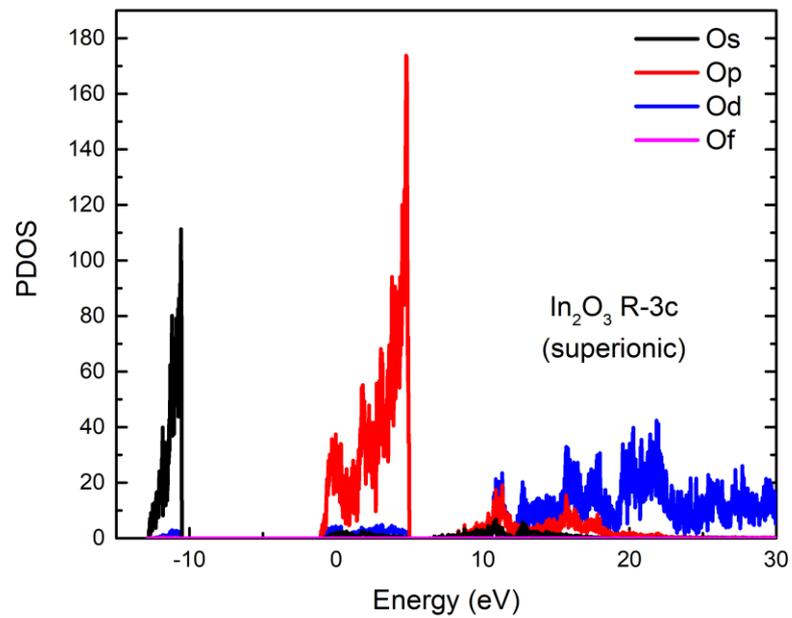

**Figure 2.2.6.** Contribution of O orbitals to electronic DOS of R-3c phase of In$_2$O$_3$.

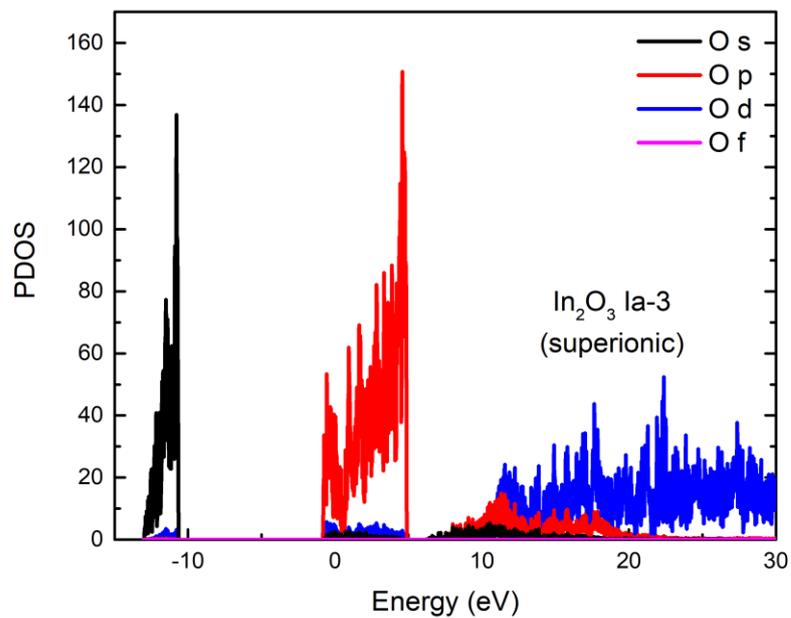

**Figure 2.2.7.** Contribution of O orbitals to electronic DOS of Ia-3 phase of $In_2O_3$.

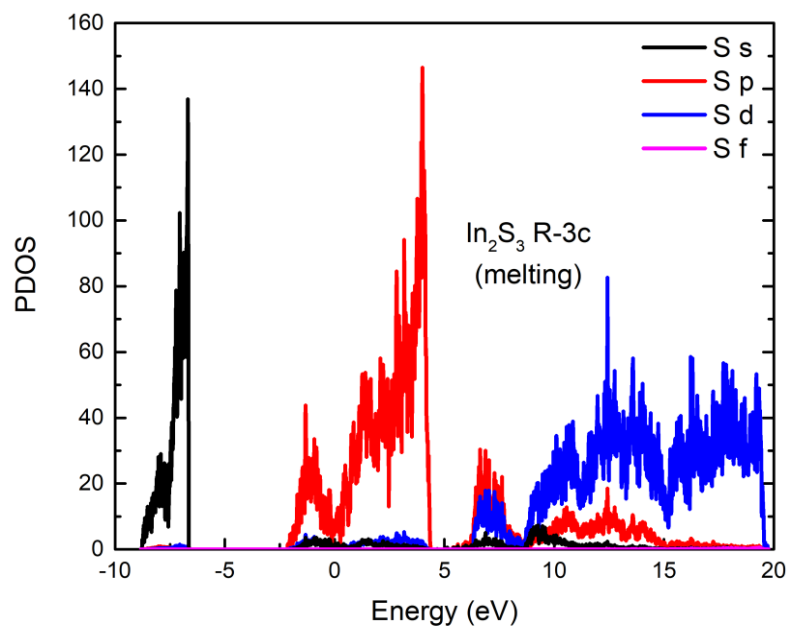

**Figure 2.2.8.** Contribution of S orbitals to electronic DOS of R-3c phase of $In_2S_3$.

# 3. Contribution of Me orbitals to electronic DOS

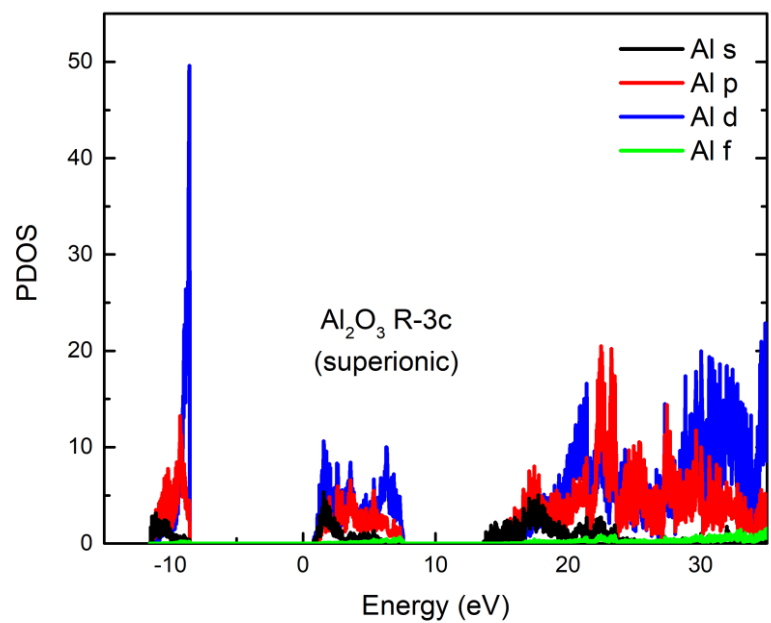

**Figure 2.3.1.** Contribution of Al orbitals to electronic DOS of R-3c phase of Al$_2$O$_3$.

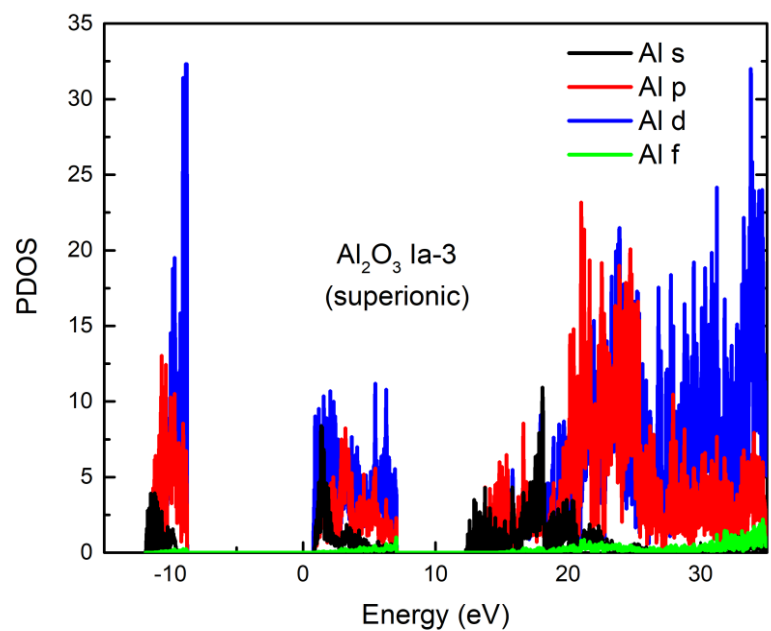

**Figure 2.3.2.** Contribution of Al orbitals to electronic DOS of Ia-3 phase of Al$_2$O$_3$.

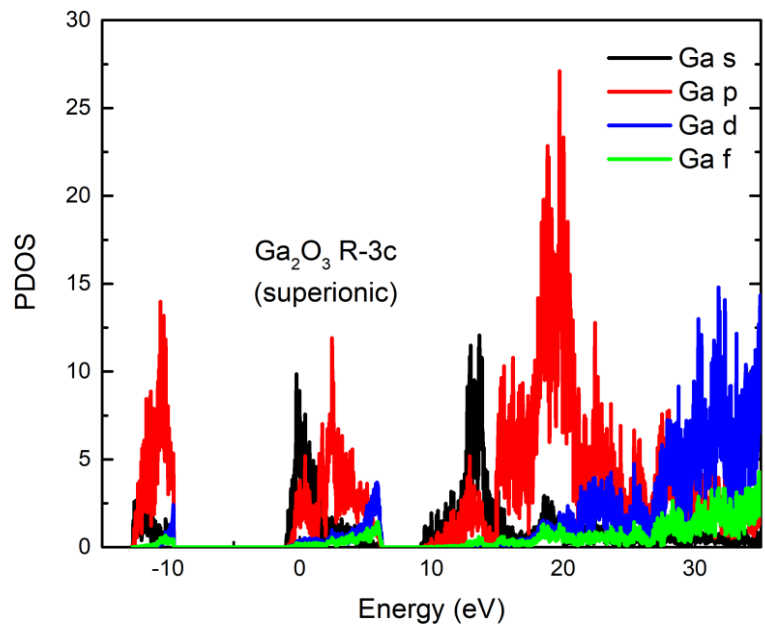

**Figure 2.3.3.** Contribution of Ga orbitals to electronic DOS of R-3c phase of Ga$_2$O$_3$.

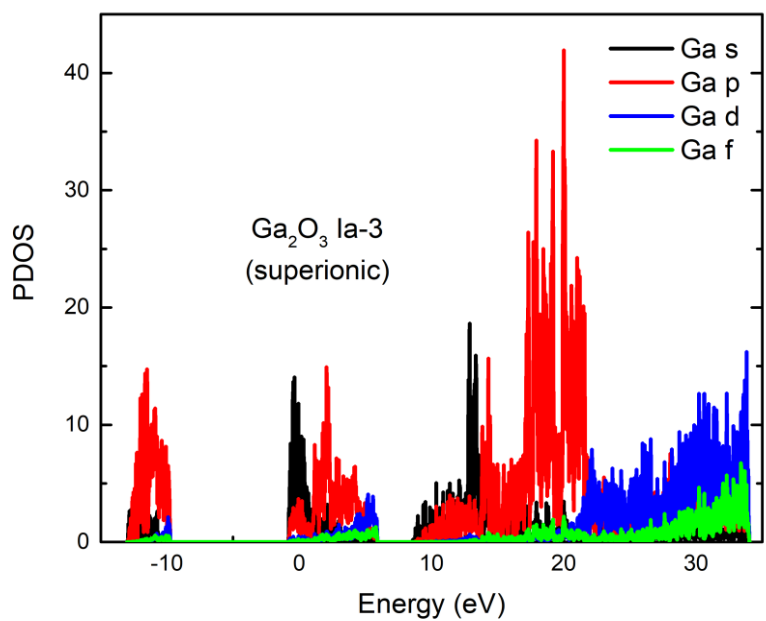

**Figure 2.3.4.** Contribution of Ga orbitals to electronic DOS of Ia-3 phase of Ga$_2$O$_3$.

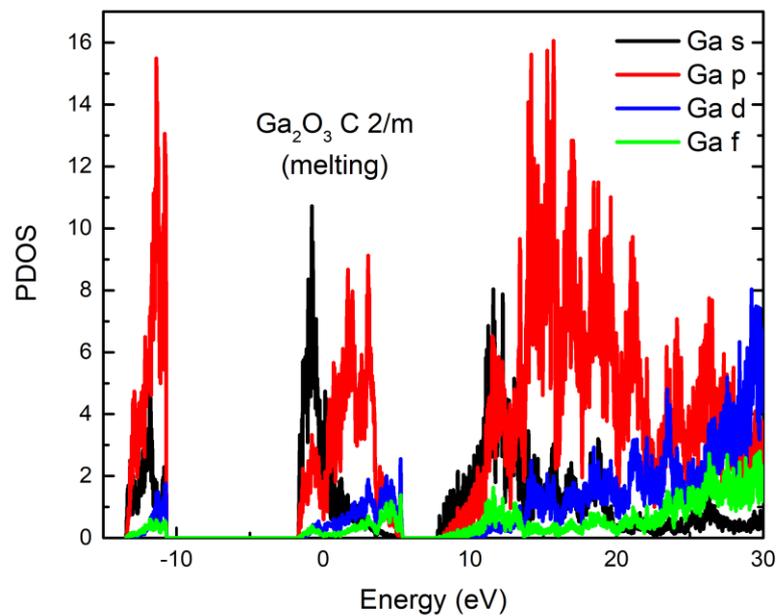

**Figure 2.3.5.** Contribution of Ga orbitals to electronic DOS of C2/m phase of Ga$_2$O$_3$.

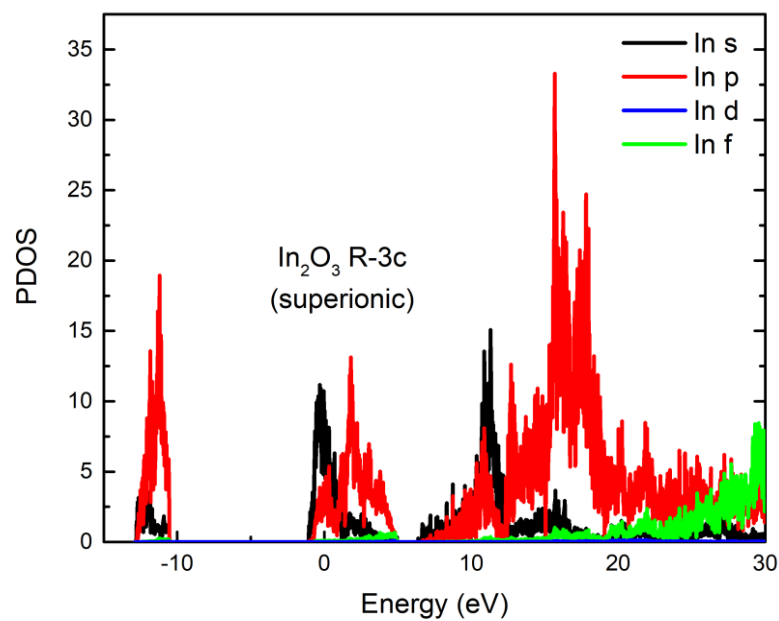

**Figure 2.3.6.** Contribution of In orbitals to electronic DOS of R-3c phase of In$_2$O$_3$.

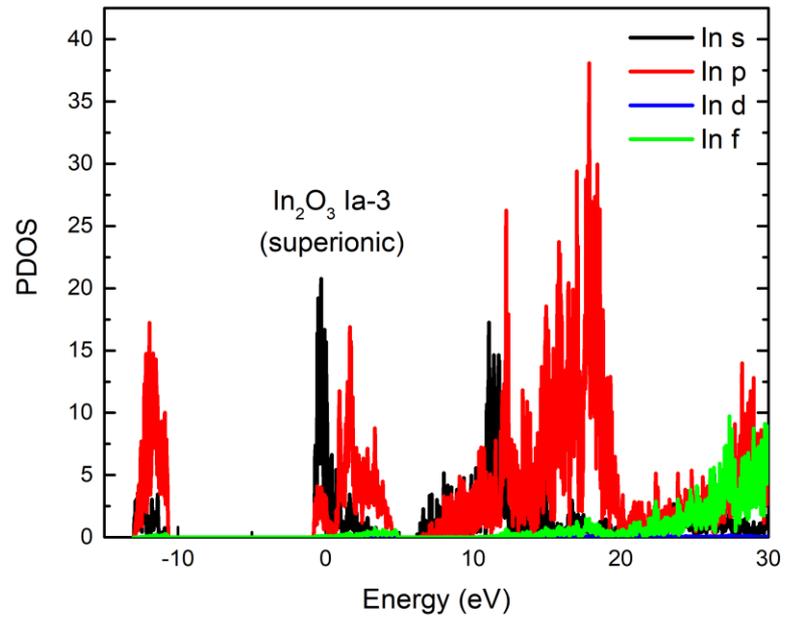

**Figure 2.3.7.** Contribution of In orbitals to electronic DOS of Ia-3 phase of In$_2$O$_3$.

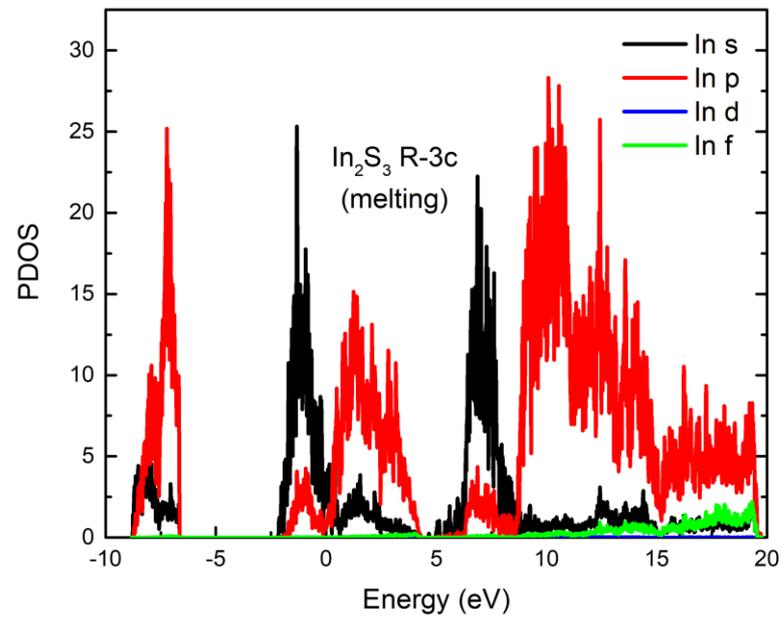

**Figure 2.3.8.** Contribution of In orbitals to electronic DOS of R-3c phase of In$_2$S$_3$.